\documentclass[12pt,titlepage,oneside,openright]{report}
\usepackage{psfig}
\title{\textbf{ \textit{On Supersymmetric Dark Matter}}}

\author{ \\ Pavel Fileviez Perez \\ \\ \\
Diploma Course in High Energy Physics. \\ The Abdus Salam International 
Centre for Theoretical Physics (ICTP). \\  Strada Costiera 11, 34100 Trieste, Italy. \\ 
and  \\ Max-Planck Institut f\"ur Physik \\ (Werner-Heisenberg-Institut) \\ 
F\"ohringer Ring 6, 80805 M\"unchen, Germany. \\ \\ \\ \\ \emph{Supervisor:} 
\\ \\  Prof. Manuel Drees \\ \\
   Physik Department, Technische Universit\"at M\"unchen, \\ James Frank Strasse, 
D-85748 Garching, Germany.}

\date{December 8, 2000}
\begin{document}

\maketitle

\abstract{The extra particles provided by Supersymmetry(SUSY) 
appear as natural candidates for exotica 
such as the missing Dark Matter of the Universe.\\
The particle candidates for Dark Matter, 
and the basic elements of the minimal supersymmetric 
extension of the Standard Model are reviewed. 
The needed elements to compute the neutralino 
counting rate are given, considering new 
Yukawa loop corrections to the 
neutralino-neutralino-Higgs 
boson couplings when the neutralino is a Bino-like state.\\
Including the new loop corrections in the Bino limit, 
the counting rate is improved in a 20\% for several values of 
the soft breaking supersymmetric parameters.}
\footnote{ Dissertation at the Diploma Course in High 
Energy Physics. ICTP. Trieste. Italy. December, 2000.}

\tableofcontents

\chapter{Introduction.}

The Dark Matter (DM) problem is one of the most interesting 
problems in modern physics. What is the Dark Matter composition? 
What is the DM energy density present in our Universe? 
What are the DM Baryonic and Non-Baryonic components? These are 
questions without a clear answer. 
\\
From several estimates it is known that the DM constitutes the dominant 
mass density in our Universe, and hence is of vital importance 
for predicting the future of the Universe.
\\ 
The DM problem establishes a very nice connection between Cosmology and 
Particle Physics, given the possibility of predicting new particle 
candidates using the well known Standard Model and its extensions. 
In the Standard Model, the possible DM candidate is the neutrino, 
while in the Minimal Supersymmetric Extension of the Standard Model (MSSM), the 
lightest supersymmetric particle (neutralino or sneutrino) appears as possible 
candidate. Many experiments are looking for DM candidates 
\cite{SDM}. If one these experiments finds DM candidates, 
it will provide a very important hint to clarify some 
fundamental ideas, such as the presence of supersymmetry at the weak scale.
\\
As the fundamental theory to break Supersymmetry (SUSY) spontaneously 
is not known, the prediction of SUSY models such as the MSSM depends 
of many free parameters introduced to break SUSY. For example in the 
MSSM the soft breaking terms are introduced and therefore we take 
many free parameters entering in the interactions and the mass 
spectrums in the model.  In the MSSM, the neutralino 
considered in many cases the ligthest supersymmetric particle is defined 
as the combination of the neutral gauginos. 
Different limits can be obtained for this particle, depending 
of the relation between the soft breaking parameters 
entering in its mass matrix. In our case we will 
consider the neutralino in the Bino-limit as the Dark Mattter 
particle candidate.
\\
In the present dissertation we will review the different particle candidates 
for Non-Baryonic Dark Matter: the Light Neutrinos, the Weakly Interacting 
Massive Particles (WIMPs) and the Axions, and the basic elements of the Minimal 
Supersymmetric Extension of the Standard Model, such as the relation 
between particles and their superpartners, the general structure of 
the MSSM lagrangian including the soft breaking terms, the 
mass spectrums of Squarks, Neutralinos and Higgs bosons, 
and the R-symmetry, which is used to define the neutralino as a stable particle.       
\\ 
In the Chapter 4 we introduce the different effective interactions to compute the 
neutralino counting rate when the neutralino is a Bino-like state. In this case 
with the objective of improve the neutralino counting rate we consider new Yukawa 
loop corrections to the neutralino-neutralino-Higgs boson couplings. Numerical results 
of the counting rate and the relic density are showed for several values of the 
soft supersymmetric parameters, where we can appreciate the increment of the counting 
rate when the new loop corrections are taking into account.
\\

\chapter{Particle Candidates for Dark Matter (DM).}

\section{Introduction.}

The phrase \textit{Dark Matter} signifies matter whose existence has been inferred
only through its gravitational effects. Evidence for DM comes from
different cosmological scales, from galatic scales of several kiloparsecs and 
clusters of galaxies (Megaparsecs) to global scales of hundreds of Mpc's.
\\
The most well known evidence of DM comes from measurements of rotational
velocities $v_{\rm{rot}}(r)$ of spiral galaxies. 
In this case at distances r greater than the extent of
the visible mass, one expects that $v_{\rm{rot}}(r) \propto r^{-1/2}$, however it remains constant. 
Therefore from this discrepancy, the existence of dark haloes is inferred to explain this effect.
\\
From different analyses in cosmology it is well known that the evolution of
our Universe depends on the matter-energy present. Usually one uses the quantity 
$\Omega= \rho / \rho_{c}$, where $ \rho $ is the average density of our 
Universe and $ \rho_{c}$ the critical density. The critical density is given by
$\rho_{c}= 1.9 \times 10^{-29} \ \rm{h^2 \ g \ cm^{-3}}$, when the Hubble constant $H_0$ is
parametrised as $h = H_0 /(100 \ \rm{km \ s^{-1} Mpc^{-1})}$, the recent measurements of the
Hubble constant give $H_0 = 73 \pm \rm{6(stat) \pm 8 (syst) \ km \ s^{-1} \ Mpc^{-1}}$, so that 
$ 0.6 \leq h \leq 0.9$.
\\
An estimate of the total matter density can be made from fluctuations in the
Cosmic Microwave Background Radiation (CMBR). The current data gives
 $\Omega_{\rm{Total}} \simeq 0.8 \pm 0.2$, in agreement with $ \Omega_{\rm{Total}} \simeq 1$ 
predicted by models of cosmic inflation \cite{Data}. 
This can be compared with the average density of visible matter in the Universe.
\\
The matter mass density may be derived from astronomical determinations
of the average mass-to-light ratio M/L for various astrophysical 
objects (see for example ref. \cite{Bo}):

\begin{equation}
  \Omega_M = \frac{M \ L}{L \ \rho_{c} }
\end{equation}
from the $M/L \sim (3-9) M_{\odot}/ L_{\odot}$ ratio measured 
in stars, is possible estimate $ \Omega_{\rm{visible}} $:

\begin{equation}
 0.002 h^{-1} \leq  \Omega_{\rm{visible}} \leq 0.006 h^{-1}
\end{equation}
with upper value $\Omega_{\rm{visible}} \simeq 0.012$

Also through measurements of X-ray emission in clusters of galaxies we can
estimate the total matter density \cite{Data}:   

\begin{equation} 
\Omega_{\rm{clusters}} \simeq 0.4 \pm 0.1
\end{equation}
On the other hand, from analyses of Big Bang nucleosynthesis, comparing the
observed abundances of $\mbox{H, He}$ and $\mbox{Li}$, and taking into
account the chemical evolution of the Universe due to stellar burning, one finds \cite{Data}:

\begin{equation}
 0.008 \leq  \Omega_{\rm{baryons}}h^2 \leq 0.04
\end{equation}
where the upper limit in this case is $  \Omega_{\rm{baryons}} \simeq 0.1$.
\\
Note that from the previous evaluation we can conclude that the major
part of matter in the Universe is not visible and  some of this Dark Matter is not
baryonic. Therefore we can define two categories: Baryonic
Dark Matter, composed of baryons which are not seen, and Non-Baryonic Dark Matter, composed of
massive neutrinos, or elementary particles which are as yet undiscovered, where the particles
which comprise non-baryonic DM  must have survived from the Big
Bang, and therefore must be stable or have lifetimes in excess of the current age of
the Universe.
\\
Also a very important point is that for purposes of galaxy formation
models, the Non-Baryonic Dark Matter is classified as \textit{hot} or \textit{cold}, depending 
on whether the DM particles were moving at relativistic or non-relativistic 
speeds at the time galaxies could just start to form.
\\
Many candidates to describe the DM are known, ranging from the axions 
with $m \leq 10^{-2} \rm{eV} = 9 \times 10^{-69} M_{\odot} $, 
to black holes of mass $ M= 10^{4} M_{ \odot}$.
\\
The main baryonic candidates are massive compact halo objects (MACHOs), 
these include brown dwarfs, Jupiters, stellar 
black holes remnants, white dwarfs, and neutron stars.
\\
As for non-baryonic candidates, we have light neutrinos as a candidate 
to describe the hot component, and to describe the cold part, 
we have axions and weakly interacting massive particles (WIMPs). 
The axion is motivated as a possible solution to the strong 
CP problem, while WIMPs are stable particles, which arise in extensions 
of the Standard Model, such as heavy fourth generation Dirac 
and Majorana neutrinos, and in SUSY models the lightest neutralino and sneutrino.
\\
In the next sections we will describe some of the particles candidates for Non-baryonic 
DM.

\section{Neutrinos.}

In units of critical density, the cosmic mass density for massive neutrinos is estimated as:

\begin{displaymath}
 \Omega_{\nu} h^{2} = \sum_{i = 1}^{3} \frac{m_{i}}{ 93 \ \rm{eV}}
\end{displaymath} 
The observed age of the Universe together with the measured expansion rate 
yields $\Omega_{DM} h^2 \leq 0.4$ so that for any of the three families \cite{R1}:

\begin{equation}
m_{\nu} \leq 40 \ \rm{eV}
\end{equation}
This limit is probably the most important astrophysical contribution 
to neutrino physics. In this case, neutrinos with a mass $ 4 \ \rm{eV} \leq m_{\nu} \leq 40 \ \rm{eV}$ 
could represent all the non-baryonic Dark Matter.
\\
When we consider the low mass neutrinos as particle DM candidates, 
two problems are present: from the perspective of structure 
formation, neutrinos as hot dark matter would require 
topological defects such as cosmic strings to form small structures such 
as galaxies. The second problem is well known from the 
phase space constraint, for example if the neutrinos are the 
DM of dwarf galaxies, $m_{\nu}$ must be few hundreds of $\rm{eV}$, which violates 
the upper limit from the overall cosmic mass density of about $40 \ \ \rm{eV} $ \cite{R1}. 
Therefore these neutrinos can not describe the haloes of dwarf galaxies. 
However there exists the possibility to consider other candidates, 
for example the neutralino \cite{SDM}.

\section{WIMPs.}

The weakly interacting massive particles (WIMPs) are particles 
with masses between $10 \rm{GeV}$ and a few TeV, and with cross section
approximately weak strength. Their relic density is approximately given by:

\begin{equation}
\Omega_{\rm{WIMP}} h^{2} \simeq \frac{0.1 \ \rm{pb} \ c}{ \langle \sigma_{a} v \rangle}
\end{equation}
where $c$ is the speed of light, $\sigma_{a}$ is the total annihilation cross 
section of a pair of WIMPs into SM particles, $v$ is the
relative velocity between the two WIMPs in their cms system, 
and $\langle \ldots \rangle$ denotes thermal averaging.
\\
Since WIMPs annihilate with very roughly weak interaction strength, it is natural to assume that their interaction 
with normal matter is also approximately of this strength. This raises the hope of detecting relic WIMPs directly, by observing
their scattering off nuclei in a detector.
\\
Perhaps the most obvious WIMP candidate is a heavy neutrino, 
however a SU(2) doublet neutrino will have a too small relic density,
if its mass exceeds a few GeV. Also, one has to require the 
neutrino to be stable. It is not obvious why a massive neutrino should
not be allowed to decay. However in SUSY models 
with exact R-parity, the lightest supersymmetric particle (LSP) is absolutely
stable. In this case we have the sneutrino and the neutralino as candidates. 
However the sneutrinos have a large annihilation cross
section, their masses would have to exceed several hundred GeV 
for them to make a good DM candidate, this would be uncomfortably 
heavy for the lightest superparticle.
\\
The neutralino still can make a good cold DM candidate \cite{NeutDM}. It is well known that for very reasonable SUSY parameters the relic 
density is $0.1 \leq \Omega_{\widetilde{\chi^0}}h^{2} \leq 0.3 $, which is the currently preferred range to the total DM density. 
\\
In the next chapter we will analyze some properties of the neutralinos, such as masses and the different physical states.

\section{Axions}

Axions are hypothetical pseudoscalar bosons, introduced to solve the strong 
CP-problem of QCD using the Peccei-Quinn Mechanism \cite{R2}
\\
The axion mass arises from its mixing with the $\pi^0$:

\begin{equation}
m_{a} \simeq 0.6 \ eV \frac{10^{7} \rm{GeV}}{f_a} 
\end{equation}
In units of the cosmic critical density one finds for the axionic mass density:

\begin{equation}
\Omega_{a}h^{2} \simeq 0.23 \times 10^{\pm 0.6}(\frac{f_{a}}{10^{12} \rm{GeV}})^{1.18} 
\overline{\theta^2_i} F(\theta_i)
\end{equation}
where $ \overline{\theta^2_i}$ is related to the average initial value of the axion 
field and $F(\theta_i)$ with $F(0)=1$ encapsules anharmonic corrections to the axion potential.
\\
With $\Omega_{a} h^2 \simeq 0.3 \times 2^{\pm 1}$ the mass of dark matter axions is found to 
be $m_a =30-1000 \mu \rm{eV}$, astrophysics limit very important to take into 
account for future experiments\cite{R2}.

\chapter{The Minimal SUSY Extension of the SM (MSSM).}

\section{Introduction.}

The Glashow-Weinberg-Salam Model or Standard Model (SM),
describes with a very good precision all electroweak processes. It is 
based on gauge invariance under the symmetry group:

\begin{equation}
G_{SM}=SU(3)_{C} \times  SU(2)_{L} \times U(1)_Y
\end{equation}  
and its partial spontaneous symmetry breaking. In table 1 we show
all its constituents, the elementary fermions (quarks and leptons), the
scalar Higgs boson, the gauge bosons, and their transformation properties
under  $G_{SM}$ (using $Q=T_3+Y/2$).
The lagrangian of the Standard Model has the following form:

\begin{equation}
\mathcal{L}_{SM}=\mathcal{L}_{fermions}+\mathcal{L}_{gauge \ bosons} +
\mathcal{L}_{scalars}+\mathcal{L}_{Yukawa}
\end{equation}
The explicit form of the SM lagrangian is well known, for our objectives
we will write explicitly only the expression of the Yukawa interaction of
the third generation:

\begin{equation}
\mathcal{L}_{Yukawa}=h_b (\bar{t} \ \bar{b})_L \Phi b_R + h_t (\bar{t} \
\bar{b})_L i
\sigma_2 \Phi^{*}t_R + h_\tau (\bar{ \nu}_\tau \ \bar{\tau})_L \Phi \tau_R
+ h.c
\end{equation}
where $h_b, h_t, h_\tau$, t, b, $\nu_\tau$ and $\tau$
are the Yukawa couplings for each quark and lepton, the top quark, the
bottom quark, the $\tau$ neutrino and the $\tau$ lepton,
respectively (with R or L chirality). Note that we can write these terms
using only one scalar field $\Phi$, which after electroweak
symmetry breaking can generate mass for all the quarks and leptons
in the SM.
\\
 As was mentioned above, the standard model has an extremely economical
Higgs sector, which accounts for all the particle masses. Baryon (B) and 
Lepton (L) numbers are automatically conserved and it is an
anomaly free Quantum Field Theory. However all is not
perfect. At present there is no unambiguous direct experimental evidence for 
the Higgs boson. Moreover there are no answers to some fundamental  
questions such as: Could we explain the origin of parity and time reversal
breaking? Could we unify the diferent gauge couplings? Could we explain
the origin of the CP violation? 
\\
Another fundamental problem in the Standard Model is in the Higgs sector. 
There is nothing to protect the mass of the Higgs boson from quadratic divergences. 
To solve this problem we can introduce a new symmetry called supersymmetry (SUSY). SUSY is a symmetry that
transforms fermions into bosons and vice versa \cite{SUSY}.
\\
Motivations for introducing SUSY are the Haag-Lopuszanski-Sohnius Theorem (SUSY is the most general symmetry of the S-matrix), 
and cancellation of the tachyon states in String Theory. However
the best (the more realistic) motivation is introduce SUSY at the weak scale to solve the problem 
related with the Higgs mass protection.
\\
Once SUSY is introduced we will have a superpartner for each particle 
in the Minimal Supersymmetry Standard Model, and one
aditional Higgs and its superpartner are introduced \cite{MSSM}. Note that if SUSY is realized
in Nature, it means that we have discovered only aproximately the fifty percent of all particles.
\\
From experiments (e.g at LEP) we know that the selectron, the 
superpartner of the electron must have a mass at least 90 GeV, more than 5 
orden of magnitude above the electron mass. Therefore
SUSY must be broken. This raises an additional fundamental problem: How can we break SUSY spontaneously? 
We know that
when a symmetry is broken spontaneously we can keep the
renormalizability and the unitarity of our model, but in this case we
don't know the fundamental theory to break SUSY spontaneously.
\\
In the MSSM, SUSY is broken explicitly by introducing \textit{soft} breaking
terms in the lagrangian. These terms are called soft because the quadratic
divergences (mass protection for the Higgs boson) are still canceled, but SUSY is
broken \cite{SUSYMD}.
\\
In the present chapter we will describe some properties of the MSSM, the
mass spectrum of the Neutralinos, Squarks and Higgs bosons, and the main
implications of the R-symmetry introduced in this model.

\begin{array}[b]{cccc}

\\ \\ & & & SU(3)_C,\ SU(2)_L, U(1)_Y \\
           \underline{ \mbox{Quarks: }} \\ \\
\pmatrix{u^i \cr d^i}_L & \pmatrix{c^i \cr s^i}_L & \pmatrix{t^i \cr
b^i}_L
& (3_C,2_L,1/3) \\ \\
 u^i_R  & d^i_R & t^i_R & (3_C,1_L,4/3) \\ \\
d^i_R & s^i_R & b^i_R & (3_C,1_L,-2/3) \\ \\
where \ i=1,2, 3 \ (colors) & & & \\ \\
            \underline{\mbox{Leptons:}} \\ \\
\pmatrix{\nu_e \cr e}_L & \pmatrix{\nu_\mu \cr \mu}_L & \pmatrix{\nu_\tau
\cr \tau}_L & (1_C,2_L,-1) \\ \\
            e_R & \mu_R & \tau_R & (1_C,1_L,-2) \\ \\

            \underline{\mbox{Scalars:}} \\ \\
           \Phi=\pmatrix{\phi^+ \cr \phi^0} & & & (1_C,2_L,1) \\ \\
            \underline{\mbox{Gauge bosons:}} \\ \\
          G^a_\mu \ with \ a=1..8 &  &  &  (8_C,1_L,0) \\ \\
            W^b_\mu \ with \ b=1..3 &  &  &  (1_C,3_L,0) \\ \\
            B_\mu &  &  &  (1_C,1_L,0) \\ \\ \\ \underline{Table \ 1. \
Standard \ Model \ Particles} & & & 
      
\end{array}

\section{Particles and their Superpartners.}

In the MSSM, we add a superpartner for each SM particles in the same
representation of the gauge group. In Table 2 we show the names of the
superpartners. For example the superpartner of the top quark is called
stop, of the photon the superpartner is the photino, and similarly for the other
particles. However, in the Higgs sector we must add another new Higgs boson
and its superpartner to write the Yukawa interactions needed to generate masses
for all quarks and to obtain an anomaly free model.
               
\begin{array}[t]{cccc}
\\ \\ \underline{ Table \ 2.\  Names \ of \ superpartners.} & & & \\ \\ \\
\underline{\rm{Fermions}} & \iff  & \underline{\rm{Sfermions}} & \\ \\
(quarks, leptons) & &  (squarks, sleptons) & \\ \\
\ s=1/2 & &  \ s=0 \\ \\
\underline{\rm{Gauge \ Bosons}} & \iff & \underline{\rm{Gauginos}} & \\ \\
(W^\pm, Z, \gamma, gluons) & &  (Wino, Zino, photino, gluinos) & \\ \\
\ s=1 & &  \ s=1/2 & \\ \\
\underline{\rm{Higgs}} & \iff &  \underline{\rm{Higgsinos}} & \\ \\
\ s=0 & &  \ s=1/2 & \\ \\
\end{array}
Usually we use the SUSY operators in the left-chiral representation,
therefore it is convenient to rewrite the SM particles (Table 1.) in the left-chiral
representation and define the superpartners accordingly.
 This leads to the superfield formalism, which makes it easier to construct SUSY invariant lagrangians \cite{JWess}
. In this case we must introduce for each
SM particle one superfield, which contains the SM particle, its
superpartner and an auxiliary unphysical field.
In Table 3 we show the third generation of the SM particles, their
superpartners and the superfields needed to write our lagrangian. Note
that we have an extended Higgs sector and the color index is omitted.

\begin{array}[t]{cccc}
       \underline{Table \ 3.\ Content \ of \ the \ MSSM.}  & & &    \\ \\
(only \ the \ third \  generation) & & & \\ \\ \\

\underline{\rm{Superfields}} & & & \\
            \underline{\rm{Vector \ Superfields}} & & & \\ \\
& \underline{\rm{Bosonic \ Fields}} & \underline{\rm{Fermionic \ Fields}} &
G_{SM} \\ \\ \\

          G^a_3 & G^a_\mu \ with \ a=1..8 & \widetilde {G^a} &
(8_C,1_L,0)
\\ \\

             G^b_2 & W^b_\mu \ with \ b=1..3 &  \widetilde{W^b}  &
(1_C,3_L,0) \\ \\

            G_1 & B_\mu &  \widetilde{B} &  (1_C,1_L,0) \\ \\

            \underline{\rm{Chiral \ Superfields}} & & & \\ \\

\underline{\rm{Leptons}} & & & \\ \\

L & \widetilde{L_L}= \pmatrix{ \widetilde{ \nu_L} \cr \widetilde{ \tau_L}} 
& L_L=\pmatrix{ \nu_L \cr \tau_L} & (1_C,2_L,-1) \\ \\

            \tau & \widetilde{ \tau^{*}_R} & \tau^C_L & (1_C,1_L,2) \\ \\

\underline{\rm{Quarks}} & & & \\ \\
  Q &  \widetilde{Q_L}= \pmatrix{ \widetilde{t_L} \cr \widetilde{b_L}} &
\pmatrix{t_L \cr b_L} & (3_C,2_L,1/3) \\ \\
            T & \widetilde{t^{*}_R} & t^C_L & (3^*_C,1_L,-4/3) \\ \\
            B & \widetilde{b^{*}_R} & b^C_L & (3^*_C,1_L,2/3) \\ \\
             
\underline{\rm{Higgs}} \\ \\

  H_1 &  \pmatrix{H^0_1 \cr H^-_1} & \pmatrix{ \widetilde{H^0_1} \cr
\widetilde{H^-_1}} & (1_C,2_L,-1) \\ \\
H_2 &  \pmatrix{H^+_2 \cr H^0_2} & \pmatrix{ \widetilde{H^+_2} \cr
\widetilde{H^0_2}} & (1_C,2_L,1) \\ \\

\end{array}

\section{The General Structure of the MSSM Lagrangian.}

We can divide the lagrangian of the minimal
extension of the Standard Model into two
fundamental parts, the SUSY invariant and
the Soft breaking term:

\begin{equation}
\mathcal{L}_{MSSM}=\mathcal{L}_{SUSY}+
\mathcal{L}_{Soft}
\end{equation}
In general we can write the SUSY invariant
term as:

\begin{equation}
\mathcal{L}_{SUSY}=\mathcal{L}_{gauge}+
\mathcal{L}_{leptons}+\mathcal{L}_{quarks}+
\mathcal{L}_{Higgs}+ \int{d^2 \theta} \ \mathcal{W} + h.c
\end{equation}
Having specified the content of the MSSM in
Table 3 (including only the third generation of matter fields), we
have to define the different terms of the
lagrangian. The term $\mathcal{L}_{gauge}$
has the following form:

\begin{equation}
\mathcal{L}_{gauge}=\frac{1}{4} \int{d^2
\theta}[2Tr(W^3W^3)+2Tr(W^2W^2)+W^1W^1]
\end{equation}
with:
\begin{equation}
 W^3_\alpha=-\frac{1}{4} \overline{D}
\  \overline{D} \exp (-G_3) \ D_\alpha
 \exp(G_3)
\end{equation}  
\begin{equation}
G_3 = \sum_{a=1}^8 \frac{ \lambda ^a}{2}
G^a_3 \\
\end{equation}

\begin{equation}
W^2_\alpha =-\frac{1}{4} \overline{D}
\ \overline{D} \exp(-G_2) \ D_\alpha
 \exp(G_2)
\end{equation}  

\begin{equation}
G_2=\sum_{b=1}^3 \frac{\sigma^a}{2} G^b_2
\end{equation}
and
\begin{equation}
W^1_\alpha=-\frac{1}{4} \overline{D}
\ \overline{D} D_\alpha G_1
\end{equation} 
where  $\lambda^a$ \ and \  $\sigma^a$
\ are the Gell-Mann and Pauli matrices, respectively. The SUSY covariant 
derivatives D and $\overline{D}$ are used in the
left chiral representation:
\begin{equation}
D_L=\frac{\partial}{\partial \theta}+
2i\sigma^\mu \overline{\theta}\partial_\mu
\end{equation}
and
\begin{equation}
\overline{D_L}=- \frac{ \partial}
{ \partial  \overline{\theta}}
\end{equation}
The lagrangians for gauge interactions of leptons, quarks and
the Higgs bosons are:
\begin{displaymath}
\mathcal{L}_{leptons}= \int{d^2 \theta
\ d^2 \overline{\theta}} \ L^{\dagger} \exp (2g_2G_2+g_1
\frac{Y_L}{2} G_1)  L+
\end{displaymath}

\begin{equation}
+ \int{d^2 \theta \ d^2 \overline{\theta}}
\ \tau^{\dagger} \exp(2g_2G_2+g_1
\frac{Y_{\tau}}{2} G_1)  \tau
\end{equation}

\begin{displaymath}
\mathcal{L}_{quarks}= \int{d^2 \theta d^2
\ \overline{\theta}} Q^{\dagger}
 \exp(2g_3G_3+ 2 g_2 G_2 + g_1 \frac{Y_Q}{2} G_1)Q+
\end{displaymath}  

\begin{displaymath}
+ \int{d^2 \theta \ d^2 \overline{\theta}}
\ T^{\dagger} \exp(g_1\frac{Y_T}{2}
G_1 - g_3 (\lambda^a)^* G^a_3) T+
\end{displaymath}
  
\begin{equation}
+\int{d^2 \theta \ d^2 \overline{\theta}}
\ B^{\dagger} \exp(g_1\frac{Y_B}{2}
G_1- g_3 (\lambda^a)^* G^a_3) B
\end{equation}  

\begin{displaymath}
\mathcal{L}_{Higgs}= \int{d^2 \theta \ d^2
\overline{\theta}} \ H^{\dagger}_1 \exp(2g_2G_2+g_1
\frac{Y_{H_1}}{2} G_1) H_1+
\end{displaymath}  

\begin{equation}
+ \int{d^2 \theta \ d^2 \overline{\theta}} \ H^{\dagger}_2
\exp(2g_2G_2+g_1 \frac{Y_{H_2}}{2} G_1) H_2
\end{equation}
where $g_3, g_2, g_1$ are the SU(3), SU(2) and U(1) coupling 
constants, respectively. The $Y_i$ represent the hypercharges of the
different superfields.
\\
We can write the superpotential as the sum of two
terms, $\mathcal{W}$=$\mathcal{W}_R$+$\mathcal{W}_{NR}$. The first conserves 
lepton (L) and baryon (B) numbers:

\begin{equation}
\mathcal{W}_R= \epsilon_{ij}
[- \mu H^i_1 H^j_2 + h_\tau H^i_1 L^j \tau
+h_b H^i_1 Q^j B+h_t H^j_2 Q^i T] 
\end{equation}  
where $\epsilon_{ij}$ is the antisymmetric tensor, 
$\mu$ the Higgs mass parameter and
$h_\tau$, $h_b$ and $h_t$ are the different
Yukawa couplings. The term $\mathcal{W}_{NR}$, which
explicitly breaks  L and B numbers, is:

\begin{equation}
\mathcal{W}_{NR}= 
\epsilon_{ij} [- \mu^{'}  H^i_2 L^j+\lambda L^i L^j
\tau + \lambda^{'} L^i Q^j B]+\lambda^{''} B B T 
\end{equation}
In $\mathcal{W}_{NR}$ the first three terms break lepton number, while the last
term breaks baryon number. If both B
and L were broken, the proton would decay very
rapidly, therefore the products $\lambda * \lambda^{''}$ and $\lambda^{'} * \lambda^{''}$ must be very
small. In the last section of the present
chapter we will analyze the R-symmetry
related with the L and B numbers conservation and 
its implications.
 \\
The soft breaking term is:

\begin{displaymath}
- \mathcal{L}_{soft}=m^2_1 \vert H_1   
\vert^2 + m^2_2 \vert H_2 \vert^2 + m^2_{12}(H_1 H_2+H^*_1
 H^*_2)+
\end{displaymath}

\begin{displaymath}
+ \widetilde{Q_L}^{\dagger} M^2_{\widetilde{Q}}
\widetilde{Q_L} + {\widetilde{t_R}}^ \dagger
m^2_ {\widetilde{t_R}} \widetilde{t_R} +
\widetilde{b_R}^\dagger m^2_ {\widetilde{b_R}}
{\widetilde{b_R}} + {\widetilde{L_L}}
^\dagger M^2_ {\widetilde{L}
}\widetilde{L_L} + {\widetilde{\tau_R}}^*
m^2_ {\widetilde{\tau_R}}
\widetilde{\tau_R}+
\end{displaymath}

\begin{displaymath}
-H_2 \widetilde{Q_L}(h_t A_t) \widetilde{t_R} 
- H_1 \widetilde{Q_L}(h_b A_b) \widetilde{b_R}
- H_1 \widetilde{L_L}(h_\tau
 A_\tau) \widetilde{\tau_R}+
\end{displaymath} 
 
\begin{equation}
+\frac{1}{2} [M_1 \overline{\widetilde{B}}   
\widetilde{B}+ M_2 \overline{\widetilde{W^b}}  
\widetilde{W^b}+ M_3 \overline{\widetilde{g^a}}
\widetilde{g^a}]
\label{l1}
\end{equation}
\\
Note that in order to describe SUSY breaking we
introduce many free parameters, and several 
terms have mass dimension less than  
4 (super-renormalizable, but not SUSY invariant).
Also the different mass terms considered lift the degeneracy between
particles and their superpartners \cite{H93}.
\section{Neutralinos.}
Once $SU(2)_{L} \times U(1)_Y$ is broken in the MSSM, fields with different
$SU(2)_{L} \times U(1)_Y$ quantum numbers can mix, if they have the same
$SU(3)_{C} \times U(1)_{em}$ quantum numbers, and the same spin. 
\\
The neutralinos are mixtures of the
$ \widetilde{B} $, the neutral $ \widetilde{W} $ and the
two neutral Higgsinos. In general these states form
four distinct Majorana fermions, which are eigenstates
of the symmetric mass matrix [in the basis ($ \widetilde{B} $, $
\widetilde{W} $, $ \widetilde{H^0_1} $, $
\widetilde{H^0_2} $)] \cite{MSSM}:
\\
\\
\begin{equation}
\small{M_0 = \left( \begin{array}{cccc}
M_1 & 0 & - M_Z \ \cos \beta \ s_{W} & M_Z \ \sin \beta \ s_W
\\ 
 0 & M_2 & M_Z \ \cos \beta \ c_W & -M_Z \ \sin \beta \ c_W  \\  
-M_Z \ \cos \beta \ s_W & M_Z \ \cos \beta \ c_W & 0 & -\mu
\\ 
M_Z \ \sin \beta \ s_W & -M_Z \ \sin \beta \ c_W  & - \mu
& 0 \end{array} \right)}
\end{equation}
\\
\\
where $s_W = \sin \theta_W$ and $c_W = \cos \theta_W$.
\\
Note that the masses of the neutralinos are determined by the values of
four parameters $ M_1$, $ M_2$, $\mu$ and $\tan \beta$ =
$\frac{v_2}{v_1}$ = $\frac{\langle H^0_2 \rangle}{\langle H^0_1 \rangle}$.
The neutralino mass term is given by:

\begin{equation}
- \mathcal{L}^{mass}_{\psi^0}= \frac{1}{2}(\psi^0)^T \ M_0 \ \psi^0 + h.c
\end{equation}
If we define the physical states as $\chi^0_i$ = $N_{ij} \psi^0_j$, the diagonal 
mass matrix is $M_D$=$N^* M_0 N^{\dagger}$. In
the limit  $M_1$, $ M_2$, $\mu \gg M_Z$, we
can diagonalize the mass matrix very easily pertubatively. In this case the
eigenstates and their masses are \cite{MD1}:
\\   
Bino-like:
\\
\begin{equation}
\widetilde{ \chi}^0_1 \simeq \widetilde{B}+ N_{13} \widetilde{H^0_1}+ N_{14}
\widetilde{H^0_2}
\end{equation}
where:
\begin{equation}
N_{13}=\frac{M_Z \sin \theta_W (M_1 \cos \beta + \mu \sin \beta)}{\mu^2 -
M^2_1}
\end{equation}
\\
\begin{equation}
N_{14}=\frac{M_Z \sin \theta_W (M_1 \sin \beta + \mu \cos \beta)}{-\mu^2 +
M^2_1}
\end{equation}
and mass:
\\
\begin{equation}
m_{\widetilde{ \chi}^0_1} \simeq M_1
\end{equation}
Wino-like:
\\
\begin{equation}
 \widetilde{\chi}^0_2 \simeq \widetilde{W}+ N_{23} \widetilde{H^0_1}+ N_{24}
\widetilde{H^0_2}
\end{equation}
with
\\
\begin{equation}
N_{23}=\frac{M_Z \cos \theta_W (M_2 \cos \beta + \mu \sin \beta)}{-\mu^2 +
M^2_2}
\end{equation}
\\
\begin{equation}
N_{24}=\frac{M_Z \cos \theta_W (M_2 \sin \beta + \mu \cos \beta)}{\mu^2 -
M^2_2}   
\end{equation}
and  mass:
\\
\begin{equation}
m_{\widetilde{\chi}^0_2} \simeq M_2
\end{equation}
Higgsino-like:
\\
\begin{equation}
\widetilde{ \chi}^0_3 \simeq  \frac{ \widetilde{H^0_1}-\widetilde{H^0_2}}{ \sqrt{2}}+ \frac{N_{31}\widetilde{B}+ N_{32}\widetilde{W}}{ \sqrt{2}}+
\frac{  N_{34} \widetilde{H^0_2}}{ \sqrt{2}}
\end{equation}
where
\\
\begin{equation}
N_{31}=\frac{M_Z \sin \theta_W  ( \sin \beta +  \cos \beta)}{(-\mu +
M_1)}
\end{equation}
\\
\begin{equation}
N_{32}=\frac{M_Z \cos \theta_W ( \sin \beta +  \cos \beta)}{(\mu -
M_2)}
\end{equation}
\\
\begin{equation}
N_{34}= - \frac{\delta_1  \cos 2 \theta_W }{1+ \sin 2 \theta}
\end{equation}
\\
and
\\
\begin{equation}
m_{\widetilde{ \chi}^0_3} \simeq \mu (1+\delta_1)
\end{equation}
\\
$\delta_1$ is:
\\
\begin{equation}
\delta_1 = \frac{M^2_Z(1+ \sin 2 \beta)}{2 \mu} \lbrack \frac{\sin^2 \theta_{W}}{(\mu
-M_1)}+ \frac{\cos^2 \theta_{W}}{(\mu -M_2)} \rbrack
\end{equation}
Higgsino-like:  
\\
\begin{equation}
\widetilde{ \chi}^0_4 \simeq \frac{\widetilde{H^0_1}+\widetilde{H^0_2}}{ \sqrt{2}} + \frac{N_{41}\widetilde{B}+ N_{42}\widetilde{W}}{\sqrt{2}}+
\frac{ N_{44} \widetilde{H^0_2}}{ \sqrt{2}}
\end{equation}  
with
\\
\begin{equation}
N_{41}=\frac{M_Z \sin \theta_W ( - \sin \beta +  \cos \beta)}{(\mu +
M_1)}
\end{equation}
\\
\begin{equation}
N_{42}=\frac{M_Z \cos \theta_W ( \sin \beta +  \cos \beta)}{(\mu +
M_2)}
\end{equation}  
\\
\begin{equation}
N_{44}=-\frac{\delta_2  \cos 2 \theta_W }{1- \sin 2 \theta}
\end{equation}
with
\\
\begin{equation}
  m_{\widetilde{ \chi}^0_4} \simeq -\mu (1+\delta_2)
\end{equation}  
$\delta_2$ is:
\\  
\begin{equation}
 \delta_2= \frac{M^2_Z(1- \sin 2 \beta)}{2 \mu} \lbrack \frac{\sin^2\theta_{W}}{( \mu
 +M_1)} + \frac{\cos^2 \theta_{W}}{( \mu +M_2 ) } \rbrack
\end{equation}
\\
Usually the unification of the gauge couplings is assumed, in this
case the parameter  $M_2 \approx 2M_1$.
Therefore from this relation we can conclude that the Wino eigenstates 
never will correpond to the LSP particle.

 Now if $M_Z \ll M_1<M_2< \mu$ the LSP will be a nearly pure Bino eigenstate, while
if $ M_Z \ll \mu < M_1 < M_2 $ the LSP is
a linear combination of the Higgsino states. For more details of neutralino masses 
in the MSSM see \cite{Nmass}.
\section{Squarks.}
After the electroweak symmetry breaking, several terms in the MSSM
lagrangian contribute to the squarks mass matrices. Ignoring flavor mixing
between sfermions, the mass matrices for the stop and sbottom
squarks are \cite{MSSM}:
\\
stop mass matrix:

\begin{equation}
\small{M^2_{ \widetilde{t}} = \left( \begin{array}{cc}
 & \\
 m^2_t + M^2_{ \widetilde{Q}} + m^2_Z (\frac{1}{2}- \frac{2}{3} \sin^2 \theta_{W}) \cos
2 \beta & -m_t( A_t + \mu \cot \beta) \\ \\
 -m_t ( A_t + \mu \cot \beta) &  m^2_t + m^2_{ \widetilde{t}_R}+ \frac{2}{3} \ m^2_Z
\sin^2 \theta_{W} \cos 2 \beta
 \\ & \\
  \end{array} \right)}
\end{equation} 
sbottom mass matrix:

\begin{equation}
\small{
M^2_{ \widetilde{b}} = \left( \begin{array}{cc}
  & \\
 m^2_b + M^2_{ \widetilde{Q}} - m^2_Z (\frac{1}{2}-\frac{1}{3} \sin^2 \theta_{W}) \cos
2 \beta & - m_b ( A_b + \mu \tan \beta) \\ \\
 -m_b ( A_b + \mu \tan \beta) &  m^2_b + m^2_{ \widetilde{b}_R} - \frac{1}{3} m^2_Z
\sin^2 \theta_{W} \cos 2 \beta
  \\  & \\
  \end{array} \right)}
\end{equation}
in the basis ($\widetilde{f}_L$,$\widetilde{f}_R$). The physical states
are defined as:

\begin{equation}
\widetilde{f}_1=\widetilde{f}_L \cos \theta_{ \widetilde{f}}
+ \widetilde{f}_R \sin \theta_{ \widetilde{f}}
\end{equation}
and
\begin{equation}
\widetilde{f}_2=- \widetilde{f}_L \sin \theta_{ \widetilde{f}}
+ \widetilde{f}_R \cos \theta_{ \widetilde{f}}
\end{equation}  
Note the contributions of the different soft breaking parameters in the
mass matrices.

\section{The Higgs Bosons in the MSSM.}

As we have mentioned before, the existence of the Higgs bosons is the main
motivation to introduce SUSY at the electroweak scale. In the MSSM the  
tree-level Higgs potential is given by \cite{SUSYMD}:

\begin{equation}
V_{Higgs}= m^2_{H_1} \vert H_1 \vert^2 + m^2_{H_2} \vert H_2 \vert^2 +
m^2_{12} (H_1 H_2 +h.c)+
\end{equation}

\begin{equation}
+ \frac{g^2_1 + g^2_2}{8}(\vert H_1 \vert  ^2 - \vert
H_2 \vert ^2) ^2 + \frac{g^2_2}{2}(H^*_1  H_2)^2
\end{equation}
where $m^2_{H_i}= m^2_i + \vert \mu \vert ^2$ with  $i=1, 2$
\\
Note that in this equation the strength of the quartic interactions is
determined by the gauge couplings.
\\
After electroweak symmetry breaking, three of the eight degrees of
freedom contained in the two Higgs boson doublets get eaten by the  
$W^{ \pm}$ and $Z$ gauge bosons. The five physical degrees of freedom that
remain form a neutral pseudoscalar boson $A^0$, two neutral scalar Higgs
bosons $h^0$ and $H^0$, and two charged Higgs bosons $H^+$ and $H^-$.
 The physical pseudoscalar Higgs boson $A^0$ is a linear combination of
the imaginary parts of $H^0_1$ and $H^0_2$, which have the mass matrix[ in
the basis  $( \frac{Im H^0_1}{\sqrt{2}}, \frac{Im H^0_2}{\sqrt{2}}]) $:

\begin{equation}
\small{M^2_I = \left( \begin{array}{cc}
-m^2_{12} \ \tan \beta & -m^2_{12} \\
-m^2_{12}  & -m^2_{12} \ \cot \beta
  \end{array} \right)}
\end{equation}  

\begin{equation}
m^2_{A^0}= tr M^2_I = -2 m^2_{12} \ / \sin(2 \beta)
\end{equation}
The other neutral Higgs bosons are mixtures of the real parts of    
$H^0_1$ and $H^0_2$, with tree-level mass matrix $[ \frac{Re H^0_1}{\sqrt{2}},
\frac{Re H^0_2}{ \sqrt{2}}] $:

\begin{equation}
\small{M^2_R = \left( \begin{array}{cc}
-m^2_{12} \tan \beta + m^2_Z \cos^2 \beta & m^2_{12} - \frac{1}{2} m^2_Z \sin 2 \beta
\\
m^2_{12} - \frac{1}{2} m^2_Z \sin 2 \beta & -m^2_{12} \cot \beta + m^2_Z \sin^2 \beta
  \end{array} \right)}
\end{equation}
In this case the eigenvalues are:
\begin{equation}
m^2_{H^0, h^0}= \frac{1}{2}[ m^2_{A^0}+m^2_Z \pm \sqrt{(m^2_{A^0}
+m^2_Z)^2 - 4m^2_{A^0}m^2_Z \cos^2(2 \beta)}]
\end{equation}
From this equation we can see that at tree level, the MSSM predict that $m_{h^0} \leq m_Z$, 
unfortunately in this case when we consider one-loop corrections, the mass of the
light Higgs boson is modified significantly. For example assuming that the stop masses 
do not exceed 1 TeV, $m_{h^0} \leq 130 \ \rm{GeV}$ \cite{JEllis}. In this example we can see the
importance of loop corrections in the computation of physical quantities.

\section{R-symmetry and its Implications.}

In the Standard Model  the conservation of Baryon (B) and Lepton (L)
number is automatic, this is an accidental consequence of the gauge group and matter content.
 In the MSSM, as we showed in the second section of
this chapter, we can separate the most general gauge invariant superpotential basically into two fundamental
parts, where the first term conserves B and L, while the second breaks
these symmetries.
\\
 In the MSSM, B and L conservation can be related to a new discrete symmetry, which can 
be used to classify the two kinds of contributions to the superpotential.
This symmetry is the R-parity, defined as:

\begin{equation}
R=(-1)^{3(B-L)}
\label{l2}
\end{equation}
Quark and lepton supermultiplets have $R=-1$, while the Higgs
 and gauge supermultiplets have $R= +1$. The symmetry principle in this case
will be that a term in the lagrangian is allowed only if the product of
the R parities is equal 1.
\\
The conservation of R-parity as defined in equation (\ref{l2}), together with spin conservation, also implies 
the conservation of another discrete symmetry,
defined such that it will be +1 for the SM particles
and -1 for all the sparticles:

\begin{equation}
R_S=(-1)^{3(B-L)+2S}
\end{equation}
Now if we impose the conservation of $R_S$, we will have some important
phenomenological consequences:

\begin{itemize}

\item  The lightest particle with $R_S=-1$, called the lightest
supersymmetric particle (LSP), must be stable.

\item Each sparticle other than the LSP must decay into a state with an
odd number of LSPs.
\item Sparticles can only be produced in even numbers from SM  
particles.
  
\end{itemize}

\chapter{Applications of SUSY Theories to the Description of DM.}

\section{Introduction.}
 Two methods have been proposed to search for relic neutralinos \cite{SDM}. The first is the direct detection, observing their scattering off 
nuclei in a detector. The second is to look for signals for ongoing neutralino annihilation. In the second case the idea is that the neutralino
 can lose energy in collisions with nuclei and can then be trapped by the gravitational field of celestial bodies, then eventually will 
become sufficiently concentrated near the center of these bodies to annihilate with significant rate. As is well known to analyze 
these possibilities, it is very important to know very well the neutralino-nucleon scattering cross section.
\\
In this chapter, knowing the basic elements of the SUSY extension of SM, such as interactions and spectrum, and assuming the neutralino as the lightest
supersymmetric particle \cite{NeutDM}, we will compute the counting rate in the limit where the neutralino is a Bino-like state. In this case 
we will include the one-loop corrections to the neutralino-neutralino-Higgs bosons couplings.

\section{Yukawa One-Loop Corrections to the Bino-Bino-Higgs Boson Couplings.}

\begin{figure}  
\psfig{file=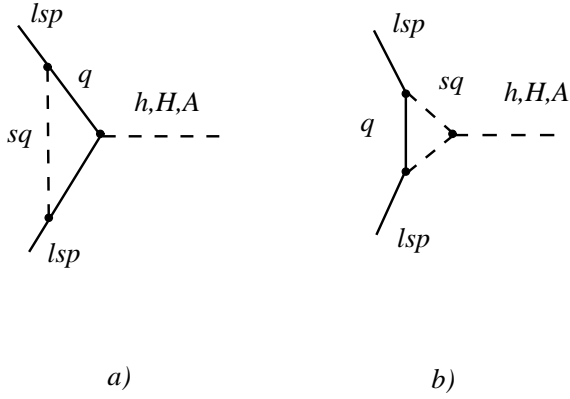,angle=0,width=0.60\textwidth}
\caption{This figure shows the new loop corrections computed, to the coupling LSP-LSP-Higgs Bosons.}
\label{f1}
\end{figure}

As we mentioned in the previous chapter, we can obtain different limits for the neutralino physical states. One of the more
interesting limits is the bino limit. In this case it is possible to compute with good precision the dark matter density contribution
needed to describe the dark haloes in the spiral galaxies.
\\
In this section we will analyze the LSP-LSP-Higgs bosons couplings in the Bino limit
at one-loop level, with the objective of computing the counting rate in direct experiments.
\\
 At tree level these couplings vanish in the pure Bino limit, therefore we will consider the one-loop correction to the elastic cross section.
\\
Using the Feynman Rules \cite{FRules}, the tree-level couplings are given by:

\begin{equation}
 i\Gamma^{tree}_{H^0 \widetilde{\chi}^0 \widetilde{\chi}^0}=i(\sin \alpha \ N_{14} - \cos \alpha \ N_{13})(g_2 \ N_{12} - g_1
\ N_{11})
\end{equation} 

\begin{equation}
i\Gamma^{tree}_{h^0 \widetilde{\chi}^0 \widetilde{\chi}^0}=i(\sin \alpha \ N_{13} + \cos \alpha \ N_{14})(g_2 \ N_{12} - g_1
\ N_{11})
\end{equation}

\begin{equation}
i\Gamma^{tree}_{A^0 \widetilde{\chi}^0 \widetilde{\chi}^0}= \gamma_5 (\sin \beta \ N_{13} - \cos \beta \ N_{14})(g_2 \
N_{12} - g_1 \ N_{11})
\end{equation}
Note that for the pure Bino state (only $N_{11} \neq 0$) all these couplings are zero at tree level.
Here the Higgs mixing angle $\alpha$ can be obtained from the following expressions:

\begin{equation}
\sin 2 \alpha = - \sin 2 \beta \frac{m^2_{H^0} + m^2_{h^0}}{m^2_{H^0} - m^2_{h^0}}
\end{equation}
and

\begin{equation}
\cos 2 \alpha = - \cos 2 \beta \frac{m^2_{A^0} - m^2_{Z^0}}{m^2_{H^0} - m^2_{h^0}}
\end{equation}
At one-loop level we will consider contributions which contain Yukawa interactions in the quark-squark loops. We therefore have to include heavy quarks 
and their superpartners inside the loops (see Fig \ref{f1}). In this case the expressions for these couplings are given by:   
\\
- Scalar Higgs bosons $h^0$ and $H^0$:

\begin{equation}
i \delta \Gamma_{ \widetilde{ \chi^0} \widetilde{ \chi^0}(h^0, H^0)} = - \frac{3i}{8
\pi^2}[h_t \delta^{(h^0, H^0)}_t + h_b \delta^{(h^0, H^0)}_b] N^2_{11}
\end{equation}
where:
\begin{displaymath}
\delta^{(h^0, H^0)}_q = C^{(h^0, H^0)}_{q} \{ \sum_{i=1}^{2} (a^2_{ \widetilde{q_i}}
- b^2_{ \widetilde{q_i}} ) [ ( m^2_{\widetilde{q_i}} + m^2_q + m^2_{ \widetilde{\chi^0}}
)C_{0}( \widetilde{q_i} ) - 4 m^2_{ \widetilde{\chi^0}} C^+_{1}( \widetilde{q_i} ) ]
\end{displaymath}

\begin{displaymath}
+ 2 m_{\widetilde{ \chi^0}}  m_q \sum_{i=1}^{2} ( a^2_{ \widetilde{q_i}} + b^2_{ \widetilde{q_i}} ) 
[C_{0} ( \widetilde{q_i} ) - 2 C^+ _{1}( \widetilde{q_i})] \}
\end{displaymath}

\begin{equation}
+ \sum_{i, j=1}^{2} C^{(h^0, H^0)}_{\widetilde{q}ij} [ m_q ( a_{ \widetilde{q}_i}
a_{ \widetilde{q}_j} - b_{ \widetilde{q}_i} b_{
\widetilde{q}_j }) C_{0} ( \widetilde{q_i}, \widetilde{q_j}) + 2m_{ \widetilde{\chi^0}}
( a_{ \widetilde{q}_i} a_{ \widetilde{q}_j} + b_{ \widetilde{q}_i} b_{
\widetilde{q}_j}) C^+_{1} ( \widetilde{q_i}, \widetilde{q_j})]
\end{equation}
-Pseudoscalar Higgs boson $A^0$.

\begin{equation}
i \delta \Gamma_{ \widetilde{ \chi^0} \widetilde{ \chi^0} A^0} =
- \frac{3i \gamma _5}{8 \pi^2} [ h_t \delta^{A^0}_t + h_b \delta^{A^0}_b ]
N^2_{11}
\end{equation}  
with:

\begin{displaymath}
\delta^{A^0}_q = C^{A^0}_{q} \{ \sum_{i=1}^{2} (a^2_{ \widetilde{q_i}} -
b^2_{ \widetilde{q_i}} ) ( - m^2_{\widetilde{q_i}} + m^2_q + m^2_{\widetilde{ \chi^0}}
) C_{0}( \widetilde{q_i} ) + 2 m_{\widetilde{ \chi^0}}  m_q \sum_{i=1}^{2}
( a^2_{ \widetilde{q_i}} + b^2_{ \widetilde{q_i}} ) C_{0} ( \widetilde{q_i} )  \}
\end{displaymath}

\begin{equation}
+ \sum_{i \neq j=1}^{2} C^{A^0}_{ \widetilde{q}ij} [ m_q ( a_{ \widetilde{q}_j}
b_{ \widetilde{q}_i} - a_{ \widetilde{q}_i} b_{
\widetilde{q}_j } ) C_{0} ( \widetilde{q_i}, \widetilde{q_j}) - 2m_{\widetilde{ \chi^0}}
( a_{ \widetilde{q}_i} b_{ \widetilde{q}_j} + a_{ \widetilde{q}_j} b_{
\widetilde{q}_i}) C^{-}_{1} ( \widetilde{q_i}, \widetilde{q_j})]
\end{equation}
Here $C^{(h^{0}, H^{0}, A^{0})}_{q}$, are the quark-quark-Higgs bosons couplings, while $C^{(h^{0}, H^{0}, A^0)}_{
\widetilde{q}ij}$ represents the $\widetilde{q_i} - \widetilde{q_j} - \mbox{Higgs bosons}$ couplings: i, j labels the squark mass eigenstates.
\\
The interaction quark-squark-neutralino is written as:

\begin{equation}
\mathcal{L}_{\widetilde{\chi^0} q \widetilde{q}_{i}} = \overline{q} (a_{\widetilde{q}_i} + b_{\widetilde{q}_i} \gamma_{5} ) \widetilde{\chi^0}
 \widetilde{q}_{i} + h.c
\end{equation}
We are using the following notation for the loop integrals \cite{loop} (see appendix A):

\begin{equation}
C_{0} ( \widetilde{q_i})=C_{0} ( k_2, k_1, m_q, m_{ \widetilde{q_i}}, m_q)
\end{equation}

\begin{equation}
C_{0} ( \widetilde{q_i}, \widetilde{q_j})=C_{0} ( k_1, k_2, m_{ \widetilde{q_i}},m_q, m_{ \widetilde{q_j}})
\end{equation}

\begin{equation}
C^{+}_{1} ( \widetilde{q_i}) = C^{+}_{1} ( -k_1, k_2, m_q, m_q, m_{ \widetilde{q_i}})
\end{equation}

\begin{equation}
C^{+}_{1} ( \widetilde{q_i}, \widetilde{q_j}) = C^{+}_{1} ( k_1, -k_2, m_{ \widetilde{q_i}}, m_{ \widetilde{q_j}}, m_q)
\end{equation}

\begin{equation}
C^{-}_{1} ( \widetilde{q_i}, \widetilde{q_j}) = C^{-}_{1} ( k_1, -k_2, m_{ \widetilde{q_i}}, m_{ \widetilde{q_j}}, m_q)
\end{equation}
where $ k_i $ represent the neutralino momentum. As already noted only consider the contribution 
of heavy quarks and their superpartners in the loop (only the third generation quarks).
\\
In the case of the top quark, using the Feynman rules given in \cite{FRules} the parameters are:
\\
Yukawa coupling:
\begin{equation}
h_{t} = \frac{g_{2} m_t}{ \sqrt{2} m_{W} \sin \beta}
\end{equation}
and
\begin{equation}
a_{ \widetilde{t_1}}= \frac{g_{2} \tan \theta_{W}}{ \sqrt{2}}(- \frac{1}{6}
\cos \theta_{\widetilde{t}} + \frac{2}{3} \sin \theta_{\widetilde{t}})
\end{equation}

\begin{equation}
a_{ \widetilde{t_2}}= \frac{g_{2} \tan \theta_{W}}{\sqrt{2}}
( \frac{1}{6} \sin \theta_{\widetilde{t}} + \frac{2}{3} \cos \theta_{\widetilde{t}})   
\end{equation}

\begin{equation}
b_{ \widetilde{t_1}}= \frac{g_{2} \tan \theta_{W}}
{\sqrt{2}}(- \frac{1}{6} \cos \theta_{\widetilde{t}} - \frac{2}{3} \sin \theta_{\widetilde{t}})
\end{equation}

\begin{equation}
b_{ \widetilde{t_2}}= \frac{g_{2} \tan \theta_{W}}{\sqrt{2}}
( \frac{1}{6} \sin \theta_{\widetilde{t}} - \frac{2}{3} \cos \theta_{\widetilde{t}})
\end{equation}
the couplings for the heavy Higgs boson $ H^0$ are given by:
\\
$\underline{t-t-H^0}$
\begin{equation}
C^{H^0}_t=- \frac{\sin \alpha}{\sqrt{2}}
\end{equation}
$ \underline{\widetilde{t}- \widetilde{t}- H^0} $

\begin{displaymath}
C^{H^0}_{\widetilde{t}11} = - \sqrt{2} [ \frac{m^2_W \sin \beta \cos ( \alpha + \beta )}
{ m_t \cos^2 \theta_W } ( \frac{1}{2} \cos^2 \theta_{\widetilde{t}} - \frac{2}{3}
\sin^2 \theta_{W} \cos 2 \theta_{ \widetilde{t}})
\end{displaymath}

\begin{equation}
 +  m_t \sin \alpha - \frac {1} {2} \sin 2
\theta_{ \widetilde{t}} ( \mu \cos \alpha + A_t \sin \alpha ) ]
\end{equation}

\begin{displaymath}
C^{H^0}_{\widetilde{t}22} = - \sqrt{2} [ \frac{m^2_W \sin \beta \cos( \alpha + \beta ) }
{ m_t \cos^2 \theta_W }  ( \frac{1}{2} \sin^2 \theta_{\widetilde{t} } +
\frac{2}{3} \sin^2 \theta_{W} \cos 2 \theta_{ \widetilde{t} } )
\end{displaymath}

\begin{equation}
 +  m_t \sin \alpha + \frac{1}{2} \sin 2 \theta_{ \widetilde{t} } ( \mu \cos \alpha + A_t \sin \alpha ) ]
\end{equation}

\begin{displaymath}
C^{H^0}_{ \widetilde{t}12} = C^{H^0}_{ \widetilde{t}21} = - \sqrt{2}
[- \frac{m^2_W \sin \beta \cos( \alpha + \beta) \sin 2 \theta_{ \widetilde{t}} }{ 4 m_t \cos^2
\theta_W}
\end{displaymath}  

\begin{equation}
 - \frac{1}{2} \cos 2 \theta_{ \widetilde{t}} (\mu \cos \alpha + A_t \sin \alpha ) ]
\end{equation}
In the case of the light Higgs boson the couplings are:
\\
$\underline{t-t-h^0}$

\begin{equation}
C^{h^0}_t=- \frac{\cos \alpha}{\sqrt{2}}
\end{equation}  

$\underline{ \widetilde{t}- \widetilde{t}- h^0} $

\begin{displaymath}
C^{h^0}_{\widetilde{t}11} = - \sqrt{2} [ - \frac{m^2_W \sin \beta \sin ( \alpha + \beta )}
{ m_t \cos^2 \theta_W } ( \frac{1}{2} \cos^2 \theta_{\widetilde{t}} - \frac{2}{3}
\sin^2 \theta_{W} \cos 2 \theta_{ \widetilde{t}})
\end{displaymath}

\begin{equation}
 +  m_t \cos \alpha + \frac {1} {2} \sin 2 \theta_{ \widetilde{t}} ( \mu \sin \alpha - A_t \cos \alpha ) ]
\end{equation}

\begin{displaymath}
C^{h^0}_{\widetilde{t}22} = - \sqrt{2} [- \frac{m^2_W \sin \beta \sin( \alpha + \beta ) }
{ m_t \cos^2 \theta_W }  ( \frac{1}{2} \sin^2 \theta_{\widetilde{t} } +
\frac{2}{3} \sin^2 \theta_{W} \cos 2 \theta_{ \widetilde{t} } )
\end{displaymath}

\begin{equation}
 +  m_t \cos \alpha - \frac{1}{2} \sin 2 \theta_{ \widetilde{t} } (  \mu \sin \alpha - A_t \cos \alpha ) ]
\end{equation}

\begin{displaymath}
C^{h^0}_{ \widetilde{t}12} = C^{h^0}_{ \widetilde{t}21} = - \sqrt{2} [ \frac{m^2_W  \sin \beta
\sin( \alpha + \beta) \sin 2 \theta_{ \widetilde{t}} }{ 4 m_t \cos^2 \theta_W}
\end{displaymath}

\begin{equation}
 + \frac{1}{2} \cos2 \theta_{ \widetilde{t}} ( \mu \sin \alpha - A_t \cos \alpha ) ]
\end{equation}
For the contribution of the pseudoscalar Higgs boson $A^0$, the couplings are:
\\
$\underline{t-t-A^0}$

\begin{equation}
C^{A^0}_t= -i  \frac{\cos \beta}{\sqrt{2}}
\end{equation}  
$\underline{ \widetilde{t} - \widetilde{t}-A^0} $

\begin{equation}
C^{A^0}_{\widetilde{t}12} = - C^{A^0}_{\widetilde{t}12}=i \frac{\sin \beta}{ \sqrt{2}} (\mu - A_t \cot \beta ) 
\end{equation}

\begin{equation}
C^{A^0}_{ \widetilde{t}11} = C^{A^0}_{ \widetilde{t}22} =0
\end{equation}  
We now turn to the sbottom coupling.
\\
The Yukawa coupling is:

\begin{equation}
h_{b} = \frac{g_{2} m_b}{ \sqrt{2} m_{W} \cos \beta}
\end{equation}
and
  
\begin{equation}
a_{ \widetilde{b_1}}= \frac{g_{2} \tan \theta_{W}}{\sqrt{2}}(- \frac{1}{6}
\cos \theta_{\widetilde{b}} - \frac{1}{3} \sin \theta_{\widetilde{b}})
\end{equation}

\begin{equation}
a_{ \widetilde{b_2}}= \frac{g_{2} \tan \theta_{W}}{\sqrt{2}}
( \frac{1}{6} \sin \theta_{\widetilde{b}} - \frac{1}{3} \cos \theta_{\widetilde{b}})
\end{equation}  

\begin{equation}
b_{ \widetilde{b_1}}= \frac{g_{2} \tan \theta_{W}}{\sqrt{2}}
(- \frac{1}{6} \cos \theta_{\widetilde{b}} + \frac{1}{3} \sin \theta_{\widetilde{b}})
\end{equation}  

\begin{equation}
b_{ \widetilde{b_2}}= \frac{g_{2} \tan \theta_{W}}{\sqrt{2}}
( \frac{1}{6} \sin \theta_{\widetilde{b}} + \frac{1}{3} \cos \theta_{\widetilde{b}})
\end{equation}
The $b-b- H^0$ coupling is given by:

\begin{equation}
C^{H^0}_b=- \frac{ \cos \alpha}{\sqrt{2}}
\end{equation}
and the couplings  $\widetilde{b}- \widetilde{b}- H^0$ are:

\begin{displaymath}
C^{H^0}_{ \widetilde{b}11} = - \sqrt{2} [ \frac{m^2_W \cos \beta
\cos ( \alpha + \beta )}{ m_b \cos^2 \theta_W } (- \frac{1}{2} \cos^2 \theta_{\widetilde{b}} + \frac{1}{3}
\sin^2 \theta_{W} \cos 2 \theta_{ \widetilde{b}})
\end{displaymath}  

\begin{equation}
 +  m_b \cos \alpha - \frac {1} {2} \sin 2 \theta_{ \widetilde{b}} (\mu \sin \alpha + A_b \cos \alpha ) ]
\end{equation}

\begin{displaymath}
C^{H^0}_{ \widetilde{b}22} = - \sqrt{2} [ \frac{ m^2_W \cos \beta \cos( \alpha + \beta ) }
{ m_b \cos^2 \theta_W }  ( - \frac{1}{2} \sin^2 \theta_{\widetilde{b} } -
\frac{1}{3} \sin^2 \theta_{W} \cos 2 \theta_{ \widetilde{b} } )
\end{displaymath}

\begin{equation}
 +  m_b \cos \alpha + \frac{1}{2} \sin 2 \theta_{ \widetilde{b} } ( \mu \sin \alpha + A_b \cos \alpha ) ]
\end{equation}

\begin{displaymath}
C^{H^0}_{ \widetilde{b}12} = C^{H^0}_{ \widetilde{b}21} = - \sqrt{2}
[ \frac{m^2_W \cos \beta \cos( \alpha + \beta) \sin 2 \theta_{ \widetilde{b}} }{ 4 m_b \cos^2
\theta_W}
\end{displaymath}

\begin{equation}
 - \frac{1}{2} \cos2 \theta_{ \widetilde{b}} (\mu \sin \alpha + A_b \cos \alpha ) ]
\end{equation}
In the case of the light Higgs boson $h^0$ the couplings are:
\\
$\underline{b-b-h^0}$
\begin{equation}
C^{h^0}_b= \frac{\sin \alpha}{\sqrt{2}}
\end{equation}
$\underline{\widetilde{b}- \widetilde{b}-h^0}$
\begin{displaymath}
C^{h^0}_{\widetilde{b}11} =  \sqrt{2} [  \frac{m^2_W \cos \beta \sin ( \alpha + \beta )}
{ m_b \cos^2 \theta_W } ( -\frac{1}{2} \cos^2 \theta_{\widetilde{b}} + \frac{1}{3}
\sin^2 \theta_{W} \cos 2 \theta_{ \widetilde{b}})
\end{displaymath}

\begin{equation}
 +  m_b \sin \alpha + \frac {1}{2} \sin 2 \theta_{ \widetilde{b}} ( \mu \cos \alpha - A_b \sin \alpha ) ]
\end{equation}

\begin{displaymath}
C^{h^0}_{\widetilde{b}22} =  \sqrt{2} [ \frac{m^2_W \cos \beta \sin( \alpha + \beta ) }
{ m_b \cos^2 \theta_W }  (- \frac{1}{2} \sin^2 \theta_{\widetilde{b} } -
\frac{1}{3} \sin^2 \theta_{W} \cos 2 \theta_{ \widetilde{b} } )
\end{displaymath}

\begin{equation}
 +  m_b \sin \alpha - \frac{1}{2} \sin 2 \theta_{ \widetilde{b} } (  \mu \cos \alpha - A_b \sin \alpha ) ]
\end{equation}

\begin{displaymath}
C^{h^0}_{ \widetilde{b}12} = C^{h^0}_{ \widetilde{b}21} =
\sqrt{2} [ \frac{m^2_W  \cos \beta \sin( \alpha + \beta) \sin 2 \theta_{ \widetilde{b}} }{ 4 m_b \cos^2
 \theta_W}
\end{displaymath}

\begin{equation}
 + \frac{1}{2} \cos 2 \theta_{ \widetilde{b}} ( \mu \cos \alpha - A_b \sin \alpha ) ]
\end{equation}
The couplings of the pseudoscalar Higgs boson $A^0$:
\\
$\underline{b-b-A^0}$
\begin{equation}
C^{A^0}_b= i  \frac{\sin \beta}{\sqrt{2}}
\end{equation}  
$\underline{\widetilde{b}- \widetilde{b}-A^0}$
\begin{equation}
C^{A^0}_{\widetilde{b}12} = - \ C^{A^0}_{\widetilde{b}12}=i \frac{\cos \beta}{ \sqrt{2}} ( \mu + A_b \tan \beta ) 
\end{equation}  
 
\begin{equation}
C^{A^0}_{ \widetilde{b}11} = C^{A^0}_{ \widetilde{b}22} =0
\end{equation}

\section{Effective Interactions.}

\begin{figure}
\centerline{\psfig{file=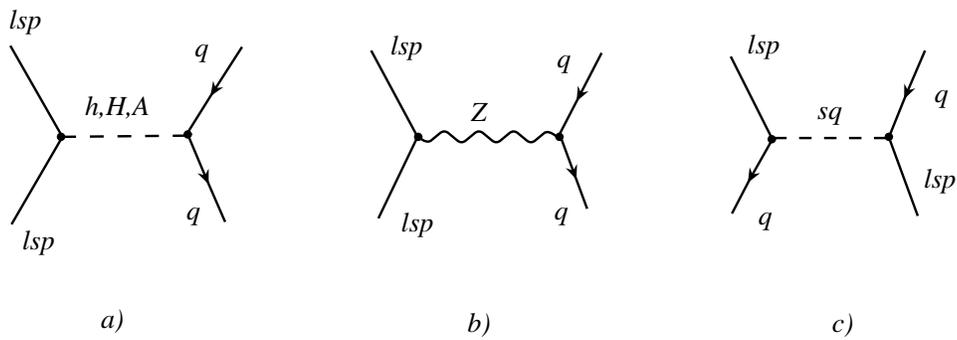,angle=0,width=0.80\textwidth}}
\caption{This figure shows the different contributions at tree level, to the neutralino-quark interaction.}
\label{f2}
\end{figure}

\begin{figure}
\centerline{\psfig{file=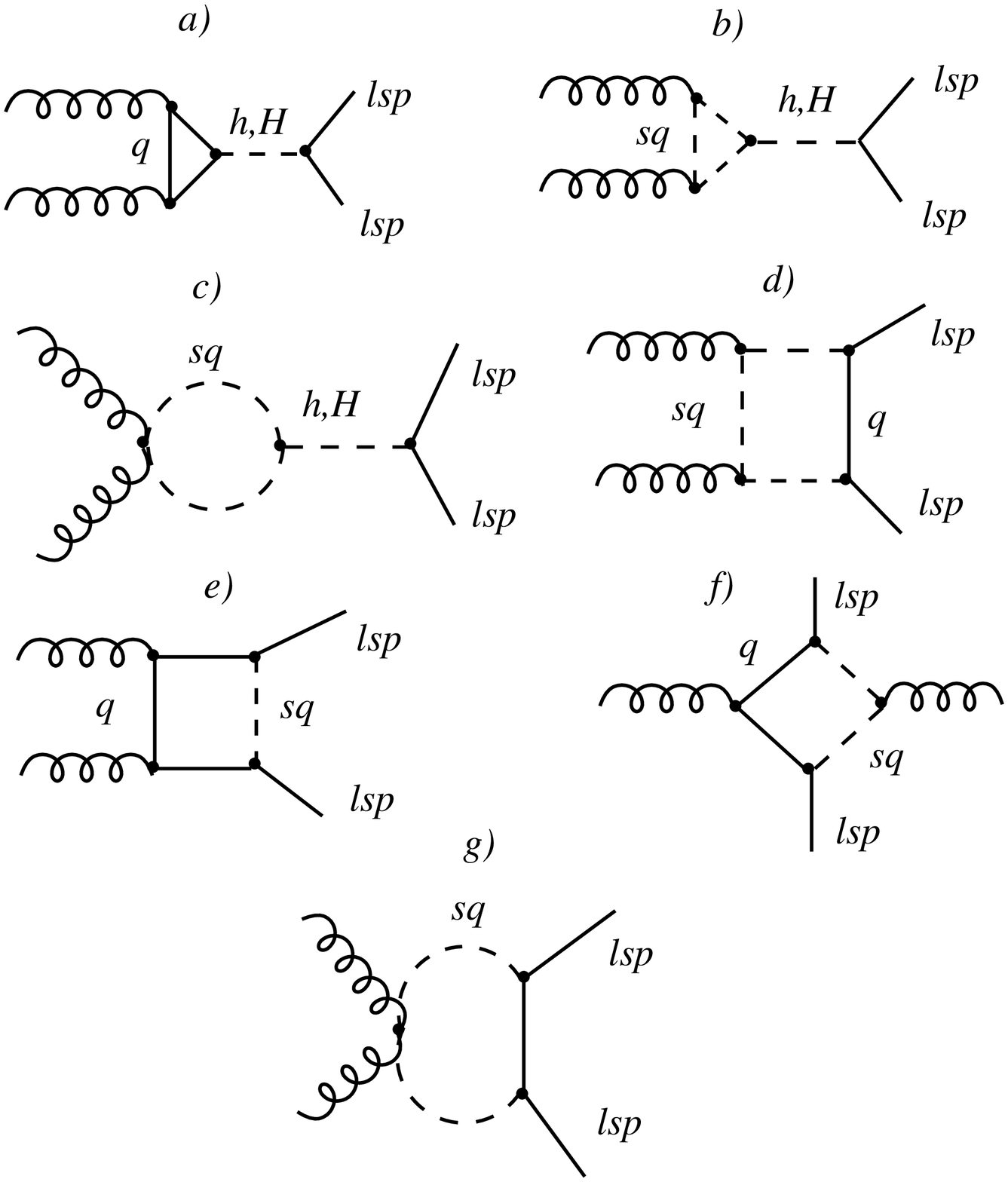,angle=0,width=0.90\textwidth}}
\caption{This figure shows the different contributions to the LSP-gluons interaction.}
\label{f3}
\end{figure}

The neutralino-nucleus elastic cross section is of fundamental importance. It determines  the
detection rate in direct and indirect detection experiments.
\\
In general the WIMP-nucleus elastic cross section depends fundamentally on the WIMP-quark interaction
strength, on the distribution of quarks in the nucleon and the distribution of nucleons 
in the nucleus \cite{SDM}.
\\
To compute the WIMP-nuclei interaction, we must first compute the interactions of WIMPs with quarks and
gluons, after using the matrix elements of the quark and gluons operators in a nucleon state we compute the
interaction with nucleons and finally using the nuclear wave functions, it is possible to compute the matrix
element for the WIMP-nucleus cross section.
\\
In the non-relativistic limit we have only two types of neutralino-quark interaction, the spin-spin interaction
and the scalar interaction. In the case of spin-spin interaction, the neutralino couples to the spin of the 
nucleus, and in the case of the scalar interaction the neutralino couples to the mass of the nucleus.
\\
To start with the analysis of the neutralino-nucleus interaction, we can discuss the effective lagrangian
describing the neutralino-quark interaction. In this case we have three classes of
contributions, the exchange of a $Z^{0}$ or Higgs boson and squark exchange (Fig \ref{f2}).
\\
From the $Z^{0}$ and squark exchange we have spin-dependent contributions, while the Higgs and
squark exchange contribute to the spin-independent interactions.
\\
The spin-dependent contribution is \cite{MD2}:
 
\begin{equation}
\mathcal{L}^{eff}_{spin} = \overline{ \widetilde{\chi^0}}   \gamma^{ \mu}   \gamma_{5}   \widetilde{\chi^0}   \overline{q}
\gamma_{ \mu} ( c_q + d_q \gamma_{5} )   q
\end{equation}
where:

\begin{equation}
c_q=- \frac{1}{2} \sum_{i=1}^{2} \frac{ a_{ \widetilde{q_i}}b_{ \widetilde{q_i}}}{m^2_{ \widetilde{q_i}}
-(m_{ \widetilde{\chi^0}} + m_q)^2}+ \frac{g^2_2}{4m^2_W} O^{R}(T_{3q}- 2 e_{q} \sin^2  \theta_W)
\end{equation}

\begin{equation}
d_q= \frac{1}{4} \sum_{i=1}^{2} \frac{ a^2_{ \widetilde{q_i}} + b^2_{ \widetilde{q_i}}}{m^2_{ \widetilde{q_i}}
-(m_{ \widetilde{\chi^0}} + m_q)^2} - \frac{g^2_2}{4m^2_W} O^{R}T_{3q}
\end{equation}
The second term in $c_{q}$ and $d_{q}$ represents the $Z^{0}$ exchange, $g_{2}$ is the SU(2) gauge
coupling, $ \theta_{W}$ the weak mixing angle, $T_{3q}$ = $ \pm 1/2 $ and $e_q$ the weak isospin and
electric charge of the quark q, and $O^{R}= \frac{1}{2}(N^2_{13}-N^2_{14})$ describes the $Z^{0} \widetilde{\chi}^0_1
\widetilde{\chi}^0_1$ coupling.
\\
To determine the spin-dependent neutralino-nucleon interaction, we must compute the matrix element:

\begin{equation}
\langle N \vert  \overline{q} \gamma_{\mu} \gamma_{5} q \vert N \rangle = 2 s_{\mu} \triangle q^{(N)}
\end{equation}
Here $s_{\mu}$ is the spin vector of the nucleon and $\triangle q^{(N)}$ denotes the second moment of the polarized quark density. 
The $ \triangle q^{(N)}$ can be extracted from
analyses of polarized deep-inelastic lepton-nucleon scattering \cite{SDM}. Here will use the values:

\begin{equation}
\triangle d^{(n)}=\triangle u^{(p)}=0.77
\end{equation}

\begin{equation}
\triangle u^{(n)}=\triangle d^{(p)}= -0.49
\end{equation}

\begin{equation}
\triangle s^{(n)}=\triangle s^{(p)}=-0.15
\end{equation}
Now knowing these values, we can write the effective spin-dependent interaction neutralino-nucleon as:
\begin{equation}
\mathcal{L}^{eff}_{spin-dep} =   \overline{ \widetilde{\chi^0}} \gamma^{\mu} \gamma_{5}  \widetilde{\chi^0} 
\overline{\Phi}_{N} s_{\mu} \Phi_{N} \sum_{q=u, d, s} 2 d_{q} \triangle q^{(N)}
\end{equation}
where $\Phi_N$ denotes the nucleon wave function. Note that this interaction is considered at tree level.
\\
The spin-independent or scalar neutralino-quark interaction is usually the most important contribution. It is given by \cite{MD2}:

\begin{equation}
\mathcal{L}^{eff}_{spin-indep} = f_q  \overline{\widetilde{ \chi^0}}    \widetilde{\chi^0} \overline{q}  q + g_q   [ -2 i O^{(2)}_{q
\mu \nu} \overline{\widetilde{ \chi^0}} \gamma^{ \mu} \partial^{ \nu} \widetilde{\chi^0} - \frac{1}{2} m_q m_{\widetilde{ \chi^0}} \overline{q} q
\overline{ \widetilde{\chi^0}} \widetilde{\chi^0} ]
\end{equation}
The leading twist-2 quark operator is:

\begin{equation}
O^{(2)}_{q \mu \nu} = \frac{i}{2} [ \overline{q} \gamma_{ \mu} \partial_{ \nu} q + \overline{q} \gamma_{
\nu} \partial_{ \mu} q - \frac{1}{2} \overline{q} \partial_{ \alpha} \gamma^{ \alpha} q g_{ \mu
\nu} ]
\end{equation}
where

\begin{equation}
 f_q=- \frac{1}{4} \sum_{i=1}^{2} \frac{ a^2_{ \widetilde{q_i}} - b^2_{ \widetilde{q_i}}}{m^2_{ \widetilde{q_i}}
-(m_{ \widetilde{\chi^0}} + m_q)^2} + m_q \sum_{j=1}^{2} \frac{ C^{(j)}_{ \widetilde{\chi}^0} C^{(j)}_q}{m^2_{H^0_j}}
\label{fq1}
\end{equation}
and

\begin{equation}
g_q=- \frac{1}{4} \sum_{i=1}^{2} \frac{ a^2_{ \widetilde{q_i}} + b^2_{ \widetilde{q_i}}}{[m^2_{ \widetilde{q_i}}
-(m_{\widetilde{ \chi^0}} + m_q)^2]^2}
\end{equation}  
The couplings  $C^{(j)}_{\widetilde{\chi}^0}$ and $C^{(j)}_q$ represent the couplings of the jth scalar Higgs boson to
the LSP and to the quark q respectively. Including the one-loop corrections obtained in the previous section, we can write $C^{(j)}_{\widetilde{ \chi^0}}$ as:

\begin{equation}
C^{(h^0, H^0, A^0)}_{\widetilde{\chi}^0} = \Gamma^{tree}_{(h^{0}, H^0, A^{0}) \widetilde{\chi}^0 \widetilde{\chi}^0} + \delta \Gamma_{(h^{0}, H^0, A^{0}) \widetilde{\chi}^0
\widetilde{\chi}^0}
\end{equation}
The different contributions to the LSP-gluons interaction are shown 
in (Fig \ref{f3}). The effective interaction from these graphs is  given by \cite{MD3}:

\begin{displaymath}
 \mathcal{L}_{gluons} = b \overline{\widetilde{ \chi^0}} \widetilde{\chi^0} G^a_{ \mu \nu} G^{a \mu \nu} -( B_{1D} + B_{1S}) \overline{\widetilde{ \chi^0}}
\partial_{ \mu} \partial_{ \nu} \widetilde{\chi^0} G^{(2) \mu \nu} +  
\end{displaymath}

\begin{equation}
 + B_{2S} \overline{\widetilde{ \chi^0}}( i \partial_{ \mu} \gamma_{ \nu} + i \partial_{ \nu} \gamma_{
\mu}) \widetilde{\chi^0} G^{(2) \mu \nu} 
\end{equation}
where:
\begin{equation}
b =  T_{ \widetilde{q}} + T_{q} + B_D +  B_S - \frac{m_{\widetilde{\chi^0}}}{2} B_{2S} - \frac{m^2_{\widetilde{\chi^0}}}{4}(B_{1D} + B_{1S})
\label{b}
\end{equation}
with
 
\begin{equation}
 B_D = \frac{g^2_3}{16 \pi^2} \frac{1}{8}\sum_{q,i} (a^2_{ \widetilde{q}_i} - b^2_{ \widetilde{q}_i}) m_q I_{1}(m_{
\widetilde{q}}, \ m_q, \ m_{ \widetilde{\chi^0}})
\end{equation}

\begin{equation}
 B_S = \frac{g^2_3}{16 \pi^2} \frac{1}{8} m_{ \widetilde{\chi^0}} \sum_{q,i} (a^2_{ \widetilde{q}_i} + b^2_{ \widetilde{q}_i})
I_{2}(m_{ \widetilde{q}},\ m_q, \ m_{ \widetilde{\chi^0}})
\end{equation}  

\begin{equation}
 B_{1D} = \frac{g^2_3}{16 \pi^2} \frac{1}{3}\sum_{q,i} (a^2_{ \widetilde{q}_i} - b^2_{ \widetilde{q}_i}) m_q
I_{3} ( m_{ \widetilde{q}},\ m_q, \ m_{ \widetilde{\chi^0}})
\end{equation}

\begin{equation}
 B_{1S} = \frac{g^2_3}{16 \pi^2} \frac{1}{3} m_{\widetilde{ \chi^0}} \sum_{q,i} (a^2_{ \widetilde{q}_i} + b^2_{ \widetilde{q}_i})  
 I_{4} ( m_{ \widetilde{q}}, \ m_q , \ m_{ \widetilde{\chi^0}})
\end{equation}

\begin{equation}
 B_{2S} = \frac{g^2_3}{16 \pi^2} \frac{1}{12} \sum_{q,i} (a^2_{ \widetilde{q}_i} + b^2_{ \widetilde{q}_i})
I_{5}(m_{ \widetilde{q}}, \ m_q, \ m_{ \widetilde{\chi^0}})  
\end{equation}

\begin{equation}
 T_{ \widetilde{q}} = \frac{g^2_3}{16 \pi}\frac{1}{12} \sum_{j=1}^{2} \frac{C^{(j)}_ {\widetilde{\chi^0}}}{m^2_{H^0_j}} \sum_{q,i} \frac{C^{(j)}_
{\widetilde{q}_i}}{m^2_{\widetilde{q}_i}}
\end{equation}

\begin{equation}
 T_{q} = \frac{g^2_3}{16 \pi}\frac{1}{3} \sum_{i=1}^{2} \frac{C^{(i)}_ {\widetilde{\chi^0}}}{m^2_{H^0_i}} \sum_{q=c, b. t} C^{(j)}_
{q}
\end{equation}
Here $ T_{ \widetilde{q}}$ and $ T_{q}$ are the Higgs contributions from squark and quark loops respectively, while all other contributions come from box and triangle
diagrams involving quarks and squarks. All coefficients with subscript D are proportional to the difference 
$ a^2_{\widetilde{q}_i} - b^2_{ \widetilde{q}_i}$, while the subscript S indicates a contribution proportional to
 $ a^2_{\widetilde{q}_i} + b^2_{ \widetilde{q}_i}$.
\\
We have used the leading twist-2 gluonic operator:
\begin{equation}
 G^{(2) \mu \nu}= G^{ a \mu }_{\rho} G^{ a \rho \nu}+ \frac{1}{4} g^{\mu \nu} G^{a \sigma \rho} G^{a}_{ \sigma \rho}
\end{equation} 
The loop integrals as function of a Feynman parameter are given by:

\begin{equation}
 I_{1}(m_{ \widetilde{q}}, m_q, m_{ \chi})= \int_{0}^{1} dx \frac{x^2-2x+2/3}{D^2}
\end{equation}

\begin{equation}
 I_{2}(m_{ \widetilde{q}}, m_q, m_{ \chi})= \int_{0}^{1} dx \frac{x(x^2-2x+2/3)}{D^2}
\end{equation}

\begin{equation}
 I_{3}(m_{ \widetilde{q}}, m_q, m_{ \chi})= \int_{0}^{1} dx \frac{x^2(1-x)^2}{D^3}
\end{equation}  
 
\begin{equation}
 I_{4}(m_{ \widetilde{q}}, m_q, m_{ \chi})= \int_{0}^{1} dx \frac{x^3(1-x)^2}{D^3}
\end{equation}

\begin{equation}
 I_{5}(m_{ \widetilde{q}}, m_q, m_{ \chi})= \int_{0}^{1} dx \frac{x(1-x)(2-x)}{D^2}
\end{equation}
with
 
\begin{equation}
 D = x^2 m^2_ {\widetilde{\chi^0}} + x (m^2_{ \widetilde{q}} -m^2_q - m^2_{ \widetilde{\chi^0}}) + m^2_q
\end{equation}
Note that this effective lagrangian is speficied at a high energy scale, for example   $\  \mu_{0} \simeq m_{\widetilde{q}}$.
\\
With the objective of writing the effective neutralino-nucleon interaction we must evaluate the matrix elements of the quark and
gluon operators in a nucleon state \cite{SDM}.
\\
For light quarks we write ($q=u, d, s$):

\begin{equation}
\langle N \vert m_{q} \overline{q} q \vert N \rangle = m_{N} f^{(N)}_{Tq} 
\end{equation}
where for the neutron we have:$ f^{(n)}_{Tu}=0.023$, $ f^{(n)}_{Td}=0.034$, $ f^{(n)}_{Ts}=0.14$, and for the proton: 
$f^{(p)}_{Tu}=0.019$, $ f^{(p)}_{Td}=0.041$ and $f^{(p)}_{Ts}=0.14$.
\\
The heavy quark matrix elements can be computed perturbatively:

\begin{equation}
\langle N \vert m_{Q} \overline{Q} Q \vert N \rangle = \frac{2}{27} m_{N} [1- \sum_{q=u, d, s}f^{(N)}_{Tq}]=
\frac{2}{27}m_{N} f_{TG}
\end{equation}
We must also compute the matrix elements of the $O^{(2)\mu \nu}$ and $G^{(2)\mu \nu} $ operators, at zero momentum transfer \cite{SDM}:

\begin{equation}
m_{N} \langle N \vert O^{(2)\mu \nu} \vert N \rangle = (p^{\mu}p^{\nu}- \frac{1}{4}m^2_N \ g^{\mu \nu}) \int^1_0 
dx \ x \ [q(x,{\mu_0}^2)+\overline{q}(x,{\mu_0}^2) ]
\end{equation}

\begin{equation}
m_{N} \langle N \vert G^{(2)\mu \nu} \vert N \rangle = (p^{\mu}p^{\nu}- \frac{1}{4}m^2_N \ g^{\mu \nu}) \int^1_0 
dx \ x \ [G(x,{\mu_0}^2) + \overline{G}(x,{\mu_0}^2) ]
\end{equation}
where $p_\mu$ is the momentum of the nucleon. The functions $q(x,{\mu}^2)$ and $G({x,\mu}^2)$ are the quark and gluon densities in
the nucleon at the scale $\mu$.
\\
Using all the matrix element mentioned before, we can write the effective lagrangian  for the scalar neutralino-nucleon interaction:

\begin{equation}
\mathcal{L}^{eff}_{scalar}= f^{(N)} \overline{\widetilde{\chi^0}} \widetilde{\chi^0} \overline{\Phi}_{N} \Phi_{N}
\end{equation}
where:

\begin{displaymath}
\frac{f^{(N)}}{m_{N}}= \sum_{q=u, d, s} \frac{f^{(N)}_{Tq}}{m_{q}}[f_q - \frac{m_{\widetilde{\chi^0}}m_q}{2}g_q]+
\frac{2}{27}f_{TG}\sum_{q=c,b,t}\frac{f^{(H)}_{q}}{m_q}-
\end{displaymath}
  
\begin{equation}
-\frac{3 m_{\widetilde{\chi^0}}}{2}\sum_{q=u, d, s, c, b} g_{q} \ q(\mu^2)-\frac{8 \pi b f_{TG}}{9} + \frac{3
m_{\widetilde{\chi^0}}}{2} \ G(\mu^2) \ g_{3}(\mu^2) \ [ B_{2S} + \frac{1}{2}m_{\widetilde{\chi^0}}(B_{1D}+B_{1S})] 
\end{equation}
Here $f^{(H)}_q $ is the Higgs boson exchange contribution to the coefficient $f_q$:

\begin{equation}
f^{(H)}_q = \sum_{i=1}^2  \frac{ g^2_2 (N_{31} \sin \alpha + N_{41} \cos \alpha)
(N_{21}- N_{11} \tan \theta_W) h_{H^{0}_{i q q}}}{4 m^2_{H^0_i}}
\end{equation} 
where $h _{H^0_i q q}$:
\begin{equation}
h_{H^0 t t}= - \frac{ m_t \sin \alpha }{m_W \sin \beta }
\end{equation}

\begin{equation}
h_{h^0 t t}= - \frac{ m_t \cos \alpha }{ m_W \sin \beta }
\end{equation}

\begin{equation}
h_{H^0 b b}= - \frac{m_b \cos \alpha }{m_W \cos \beta }
\end{equation}

\begin{equation}
h_{h^0 b b }= \frac{m_b \sin \alpha }{ m_W \cos \beta }
\end{equation}
The new loop-corrections computed appear in the Higgs exchange contribution to $f_q$ (eq \ref{fq1}), 
as well as in $f^{(H)}_q$ and the Higgs exchange contributions $T_{\widetilde{q}}$ to b (equation \ref{b}).

\section{The Counting Rate.}

The most direct possibility to detect DM particles is look for their scattering off the nuclei 
in a low-background detector. In general the counting rate is defined as events per days per kilogram of target, 
depending on the density and velocity distribution of dark matter particles in the galactic halo near 
the Earth. In their velocity distribution, one should take into account the motion of the Sun and Earth, 
which increase the total rate and give a yearly modulation to the differential rate. It will be important 
as method of distinguishing signal and noise if many events are found. More details about nuclear physics of DM can be found in \cite{NPDM} and \cite{SDM} 
\\
In general the differential detection rate can be written as \cite{MD2}:

\begin{equation}
\frac{dR}{dE_{R}} = N_{T} \frac{\rho_{\widetilde{\chi^0}}}{m_{\widetilde{\chi^0}}}
\int d \vec{v} f(\vec{v}) \frac{d \sigma}{dE_R} (v, E_{R})
\end{equation} 
where $N_{T} = \frac{6.02 \times 10^{26}}{A}$ is the number of the target 
nuclei per kg, $\rho_{\widetilde{\chi^0}} $ is the local neutralino matter 
density, $\vec{v}$ and $f(\vec{v})$ denote the velocity and the velocity distribution function 
in the Earth frame, $E_{R}= \frac{m_{r}^{2}v^{2}(1-\cos\theta^{*})}{m_N}$ is the nuclear recoil energy, where $\theta^{*}$ is the 
scattering angle in the neutralino-nucleus center of mass frame, $m_N$ is the nuclear mass and $m_r$ the neutralino-nucleus reduced mass.
\\
Considering a Maxwell distribution for the galatic neutralinos, a Gaussian nuclear form factor and 
integrating over all the possible incoming velocities and deposited energies, the counting rate is:

\begin{equation}
R= \frac{\sigma \xi}{m_{\widetilde{\chi^0}}m_N} \frac{1.8 \times 10^{11}GeV^{4}}{kg \ days} 
 \frac{\rho_{\widetilde{\chi^0}}}{0.3 \ GeV/ cm^3} \frac{\overline{u_{\widetilde{\chi^0}}}}{320 \ km/s}
\end{equation}
where $\overline{u_{ \widetilde{ \chi^0}}}$ is the average neutralino velocity, which results 
from the integration over the Maxwellian velocity distribution. $\sigma$ is the elastic 
neutralino-nucleus cross section at zero momentum transfer, it is given by:

\begin{equation}
\sigma= \frac{4 m^2_{\widetilde{\chi^0}} m^2_N}{\pi (m_{\widetilde{\chi^0}}+ m_N)}
\{[Z f^{(p)}+(A-Z)f^{(n)}]^2 + 4 \lambda^2 J(J+1)[\sum_{u, d, s} d_{q}\triangle q^{(N)}]^2 \}
\end{equation}
We will consider as target the isotope 73 of $Ge$ and $\lambda ^2 J(J+1)=0.065$ for this nucleus, where
$J$ is the total spin of the nucleus and $\lambda$ the nucleonic matrix element.

\begin{equation}
\xi = \frac{0.573}{B} \{   \frac{\exp[-B/(1+B)]}{\sqrt{1+B} } \frac{erf[1/ \sqrt{(1+B)}]}{erf(1)} \}
\end{equation}
describes the suppression due to the Gaussian nuclear form factor considered. $erf(\ldots)$ is the error function and 

\begin{equation}
B = \frac{m^2_{\widetilde{\chi^0}} m^2_N}{ (m_{\widetilde{\chi^0}}+ m_N)^2} \frac{8}{9} r^{2} \overline{v_{\widetilde{\chi^0}}}
\end{equation}
with $\overline{v_{\widetilde{\chi^0}}}= \frac{ \overline{u_{\widetilde{\chi^0}}}}{1.2}$.
\\
For the spin-independent interaction, we will use $r_{\rm{spin-indep}}=(0.3 + 0.89 A ^{1/3}) \ fm$, 
while for spin-dependent 
interaction $r_{\rm{spin-dep}}=1.25 \ r_{\rm{spin-indep}}$.
\\
\\
Before showing some numerical results we must mention different aspects taken into account: 

\begin{itemize}

\item  The neutralino is considered the lightest supersymmetric particle. It means that we must compute all the sparticle masses and impose this condition.

\item  We will take into account the experimental constraints from SUSY search experiments, given in reference \cite{Data}: 
$m_{\widetilde{\chi^+_1}} > 90 \ GeV, m_{\widetilde{e_R}} > 89 \ Ge V $ and $m_{\widetilde{\nu}} > 43 \ Ge V $. The 
experimental contraints on the Higgs search also play an important role.

\item Also we will show the values for the neutralino density, to estimate the range where $\Omega h^2$ has an interesting value for cosmology.
\\
The relic density is computed, taking into account the annihilation of neutralinos in: 
two scalar neutral Higgs bosons, one scalar plus one pseudoscalar Higgs boson, $f$ plus $\overline{f}$, 
with $f$ being an Standard Model fermion, $Z$ plus Higgs bosons, $W^- \ W ^+$, $Z \ Z$, $W^+ H^-$, $H^- H ^+$ and the 
co-annihilation with charginos and sfermions.
\end{itemize}

\begin{figure}
\centerline{\psfig{file=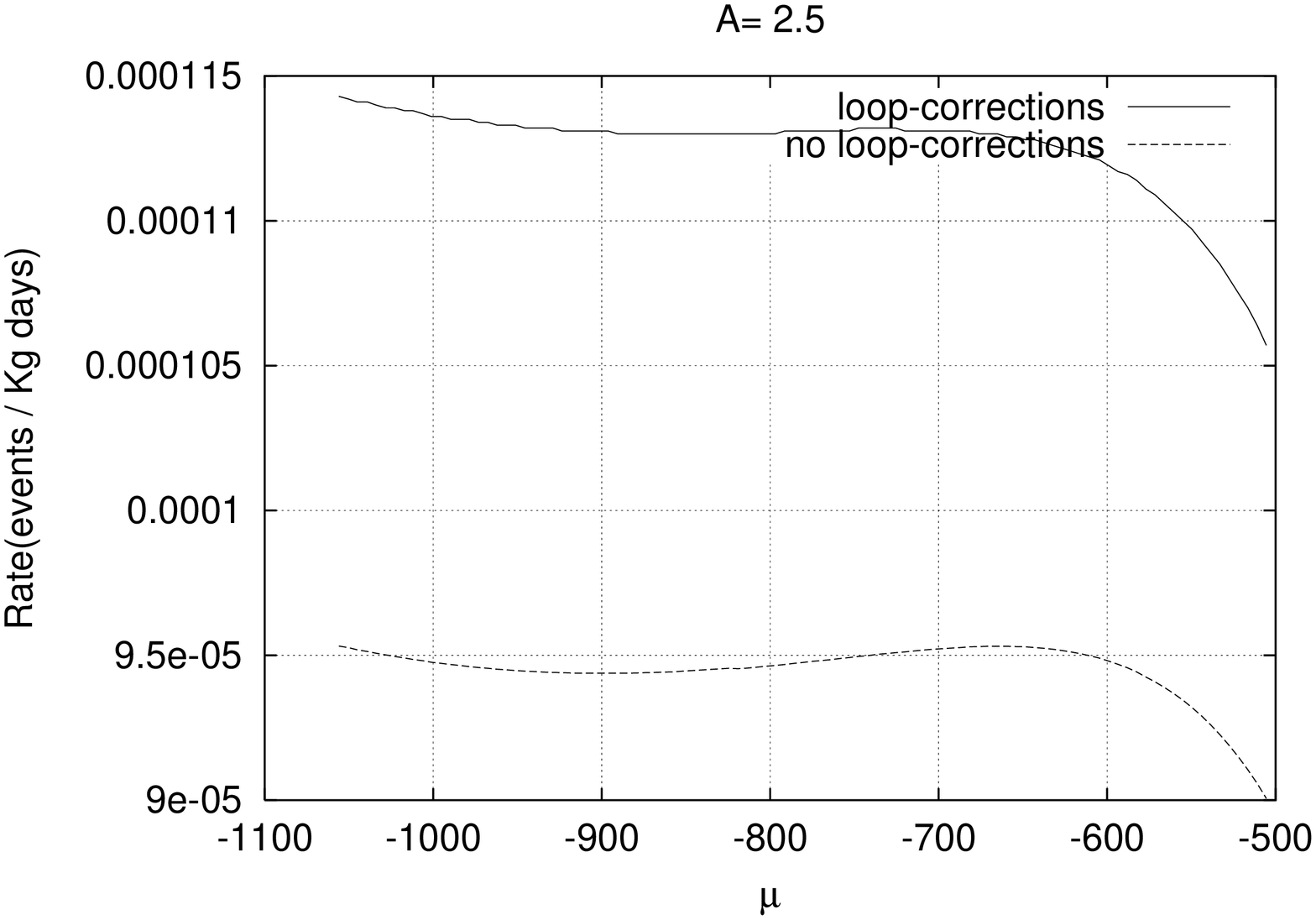,angle=-90,width=0.60\textwidth}}
\caption{}
\label{f4}
\end{figure}

\begin{figure}
\centerline{\psfig{file=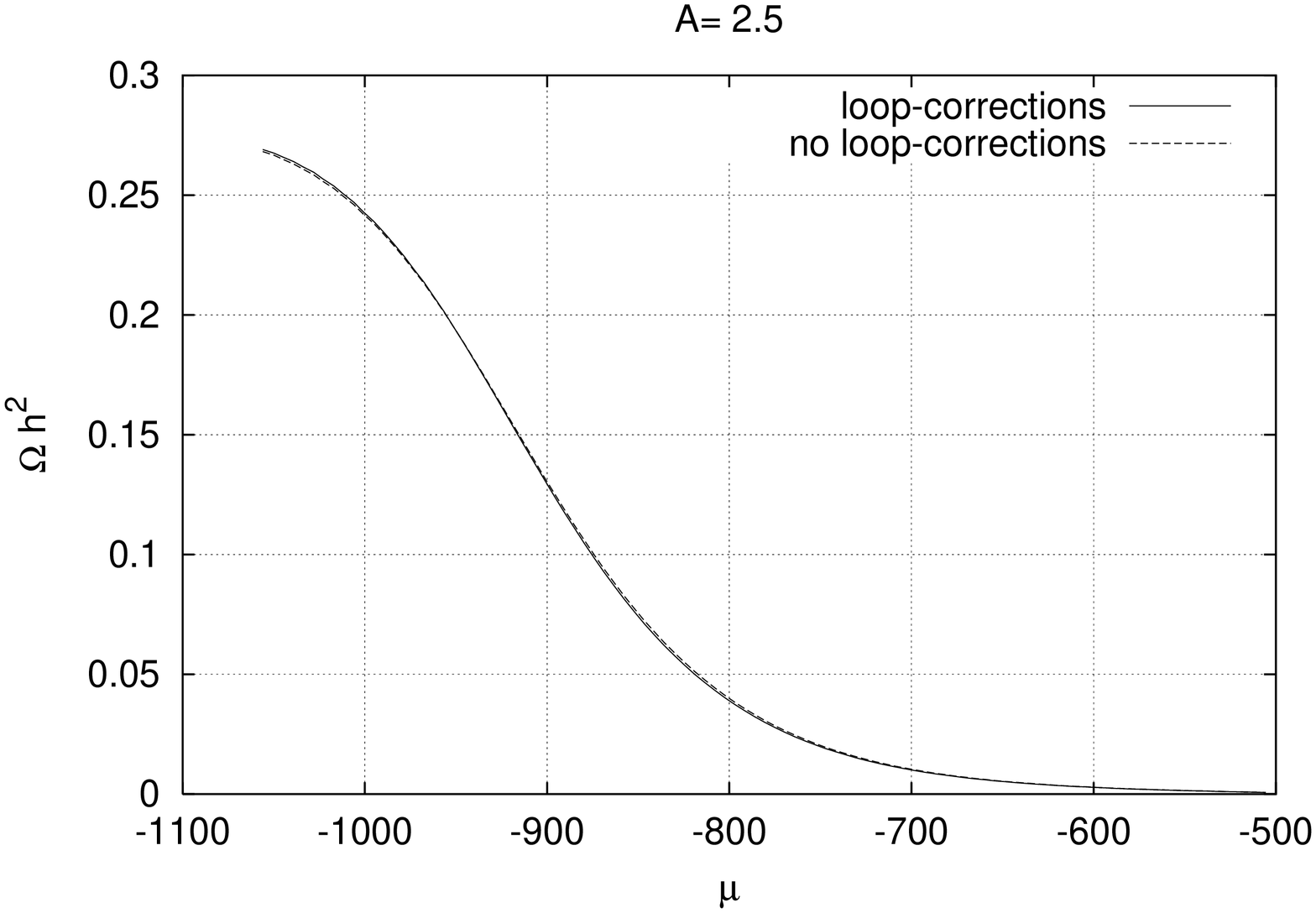,angle=-90,width=0.60\textwidth}}
\caption{}
\label{f5}
\end{figure}

\begin{figure}
\centerline{\psfig{file=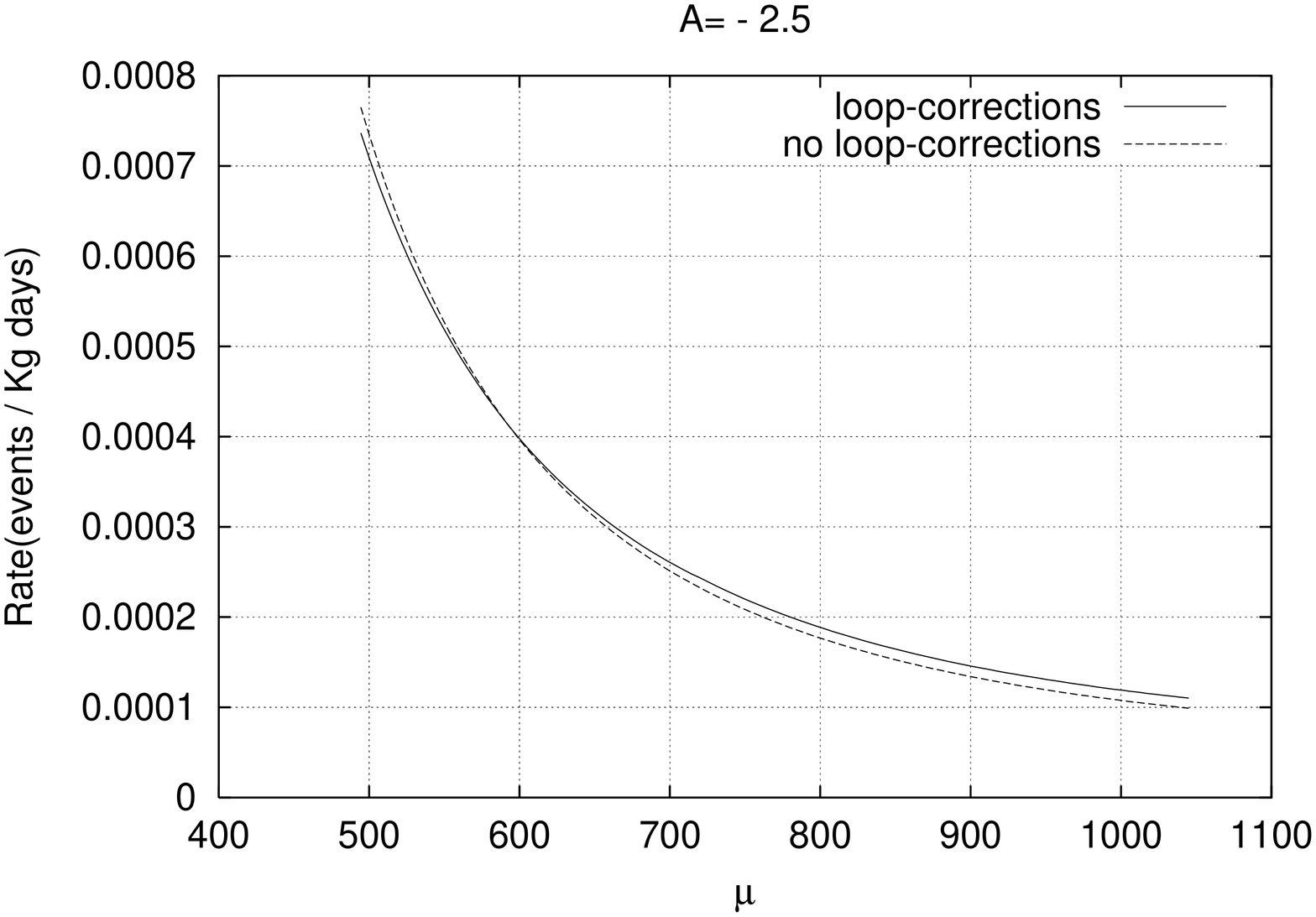,angle=-90,width=0.60\textwidth}}
\caption{}
\label{f6}
\end{figure}

\begin{figure}
\centerline{\psfig{file=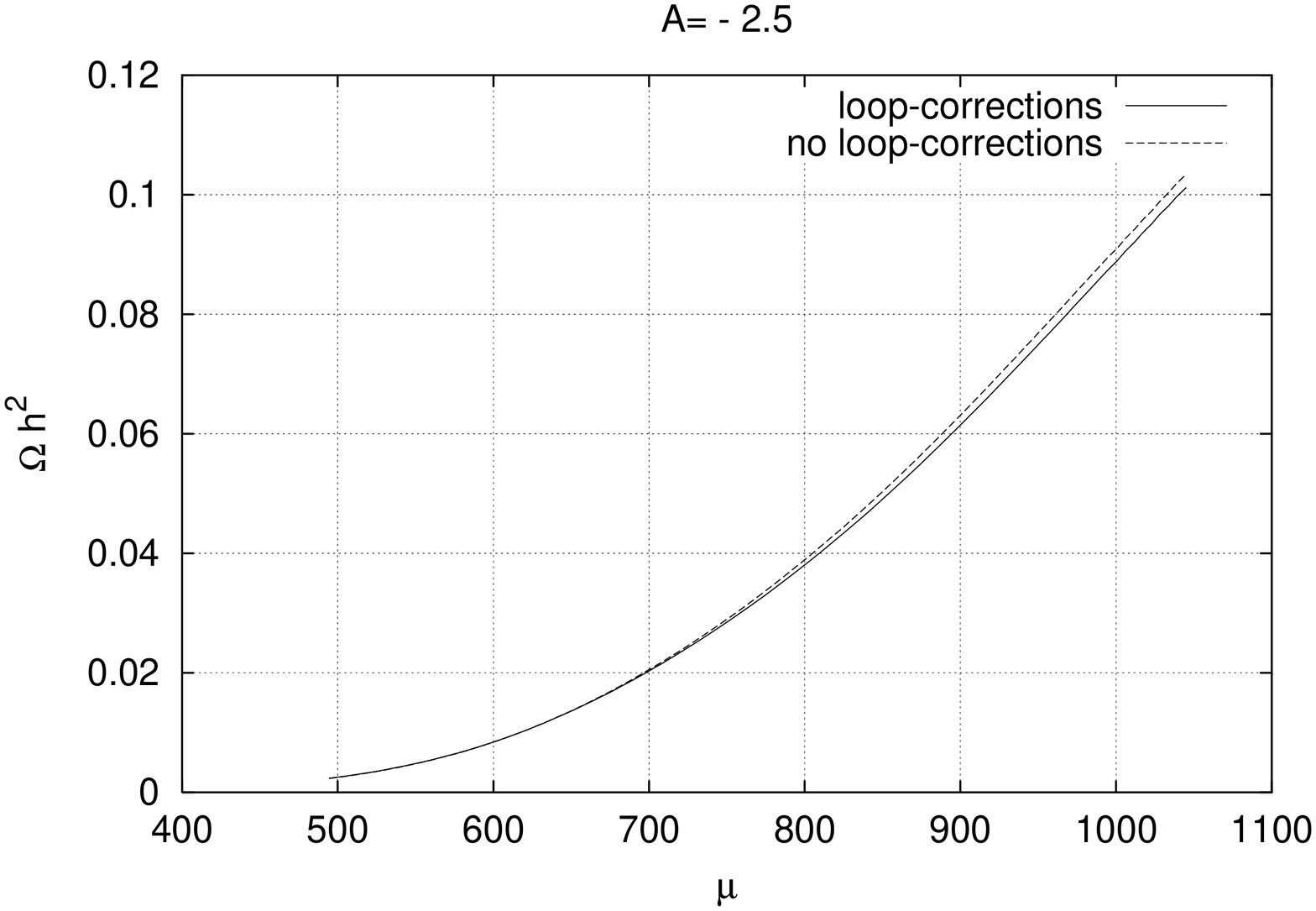,angle=-90,width=0.60\textwidth}}
\caption{}
\label{f7}
\end{figure}

\begin{figure}
\centerline{\psfig{file=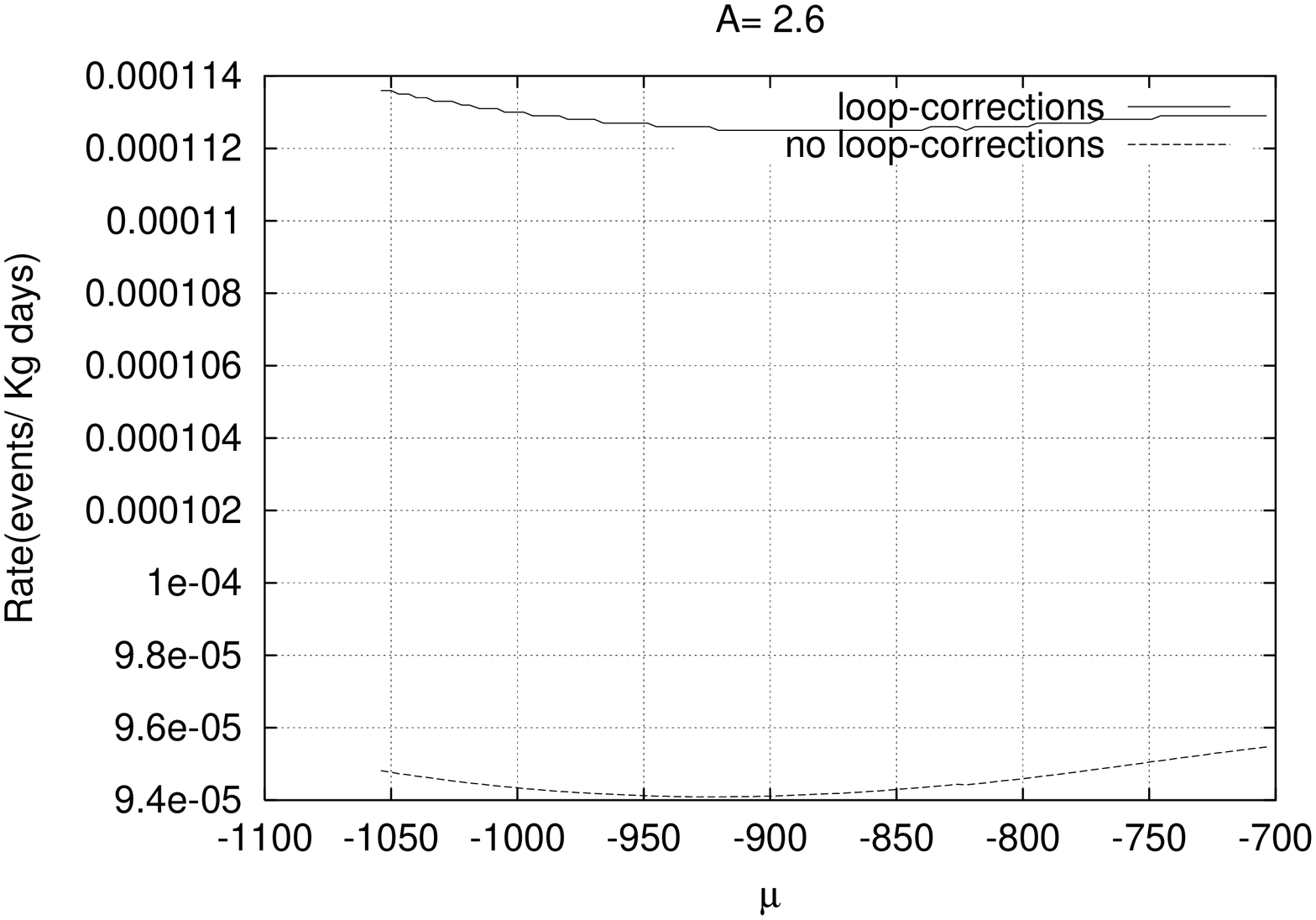,angle=-90,width=0.60\textwidth}}
\caption{}
\label{f9}
\end{figure}

\begin{figure}
\centerline{\psfig{file=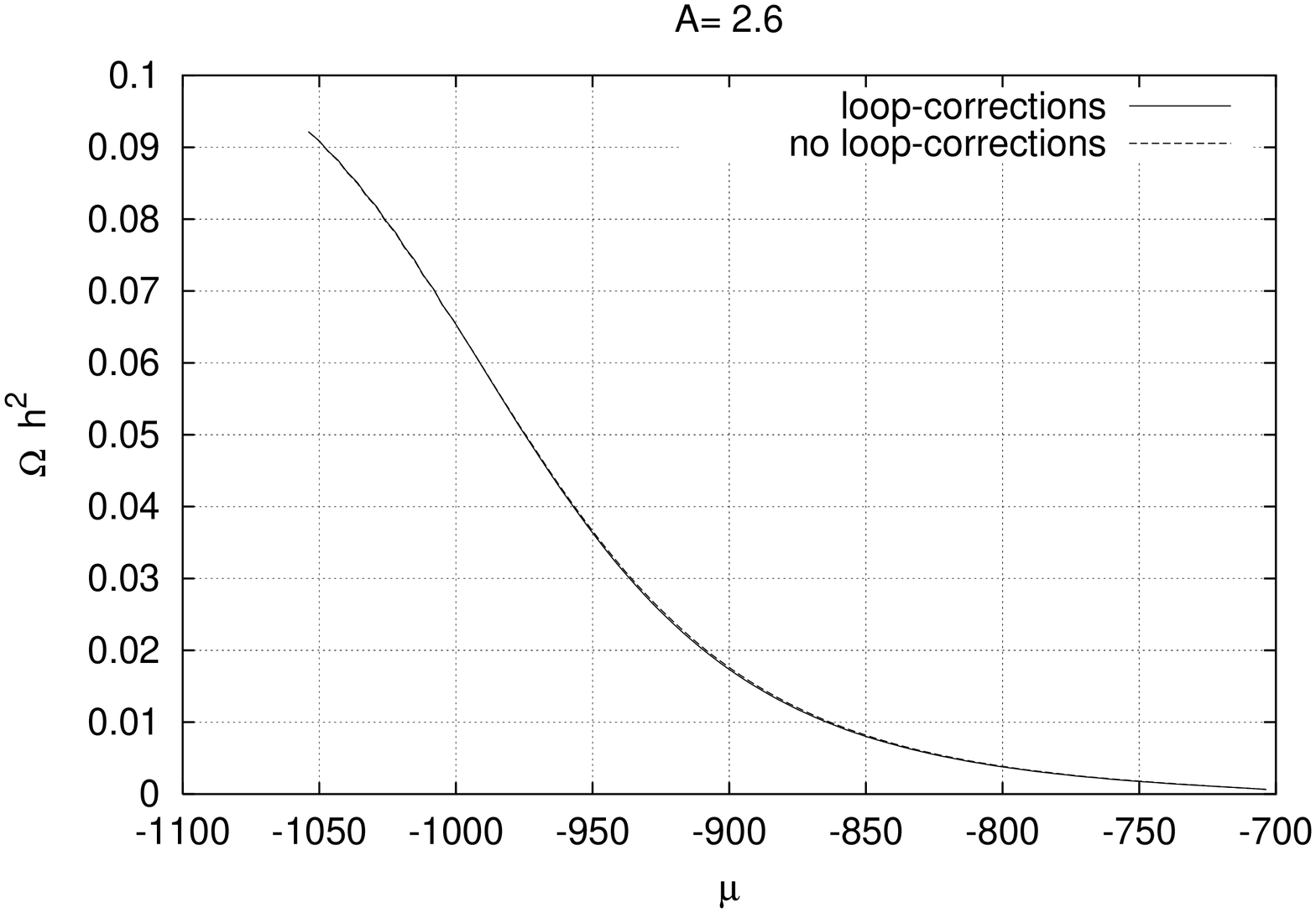,angle=-90,width=0.60\textwidth}}
\caption{}
\label{f10}
\end{figure}

\begin{figure}
\centerline{\psfig{file=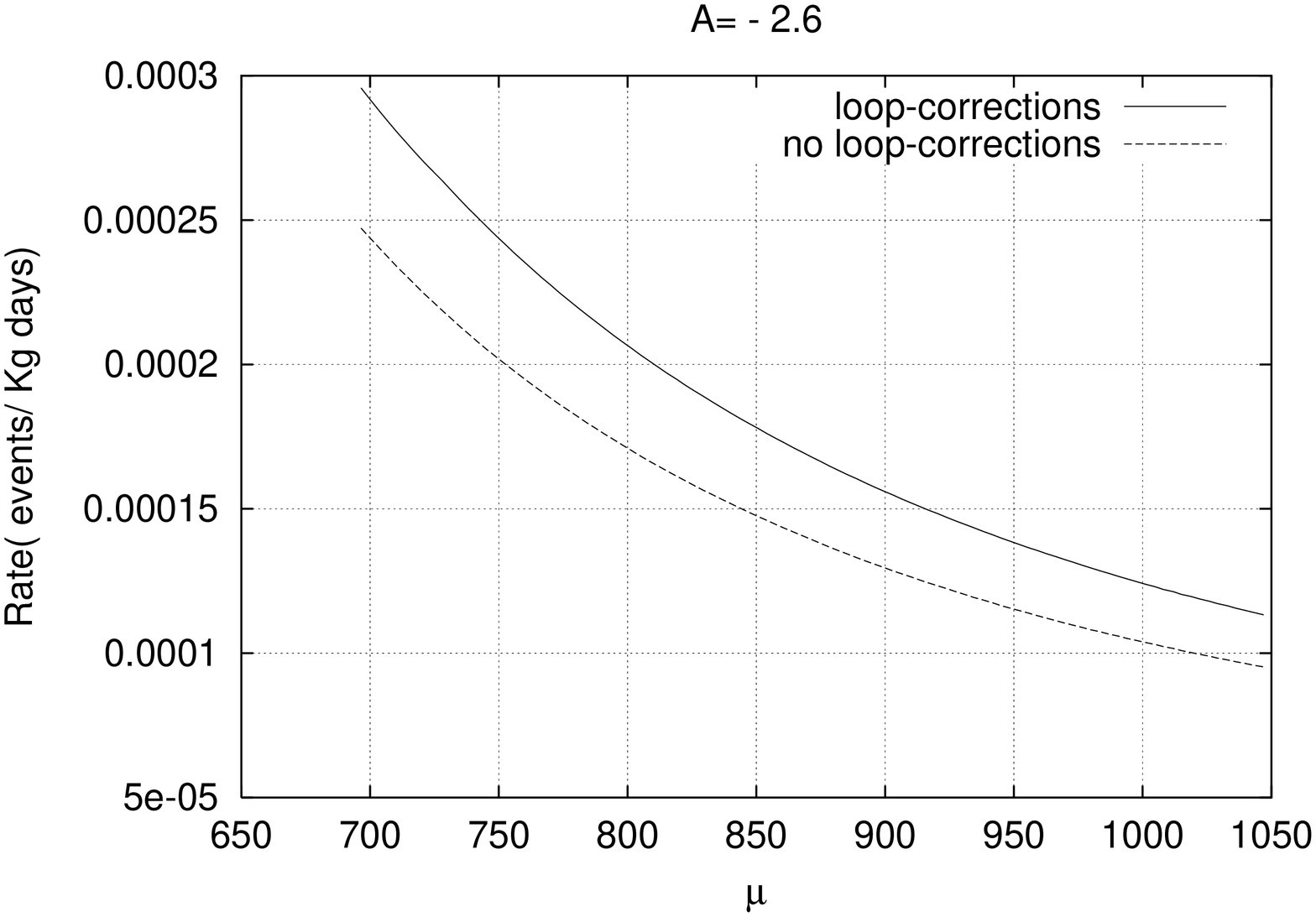,angle=-90,width=0.60\textwidth}}
\caption{}
\label{f11}
\end{figure}

\begin{figure}
\centerline{\psfig{file=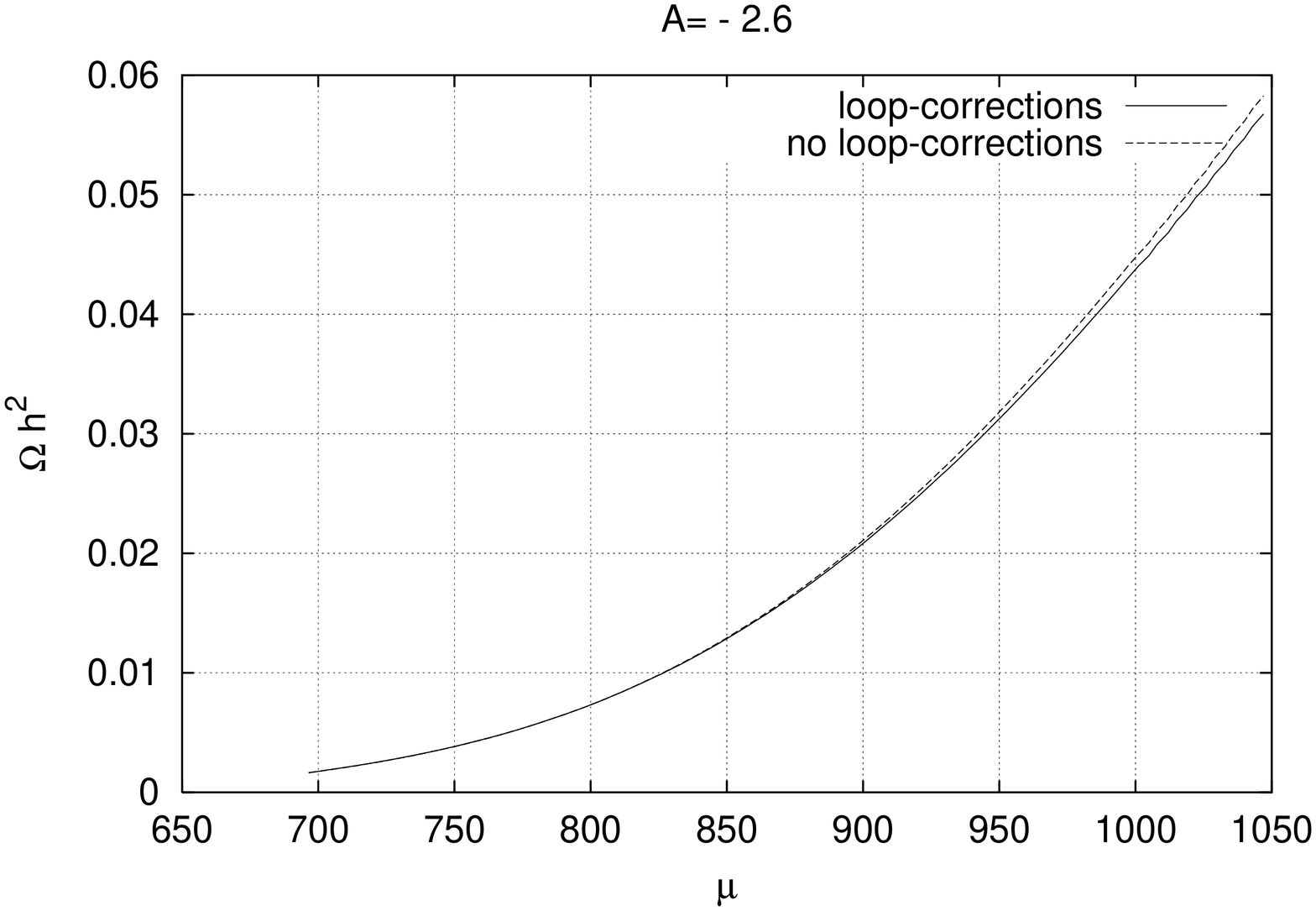,angle=-90,width=0.60\textwidth}}
\caption{}
\label{f12}
\end{figure}

\begin{figure}
\centerline{\psfig{file=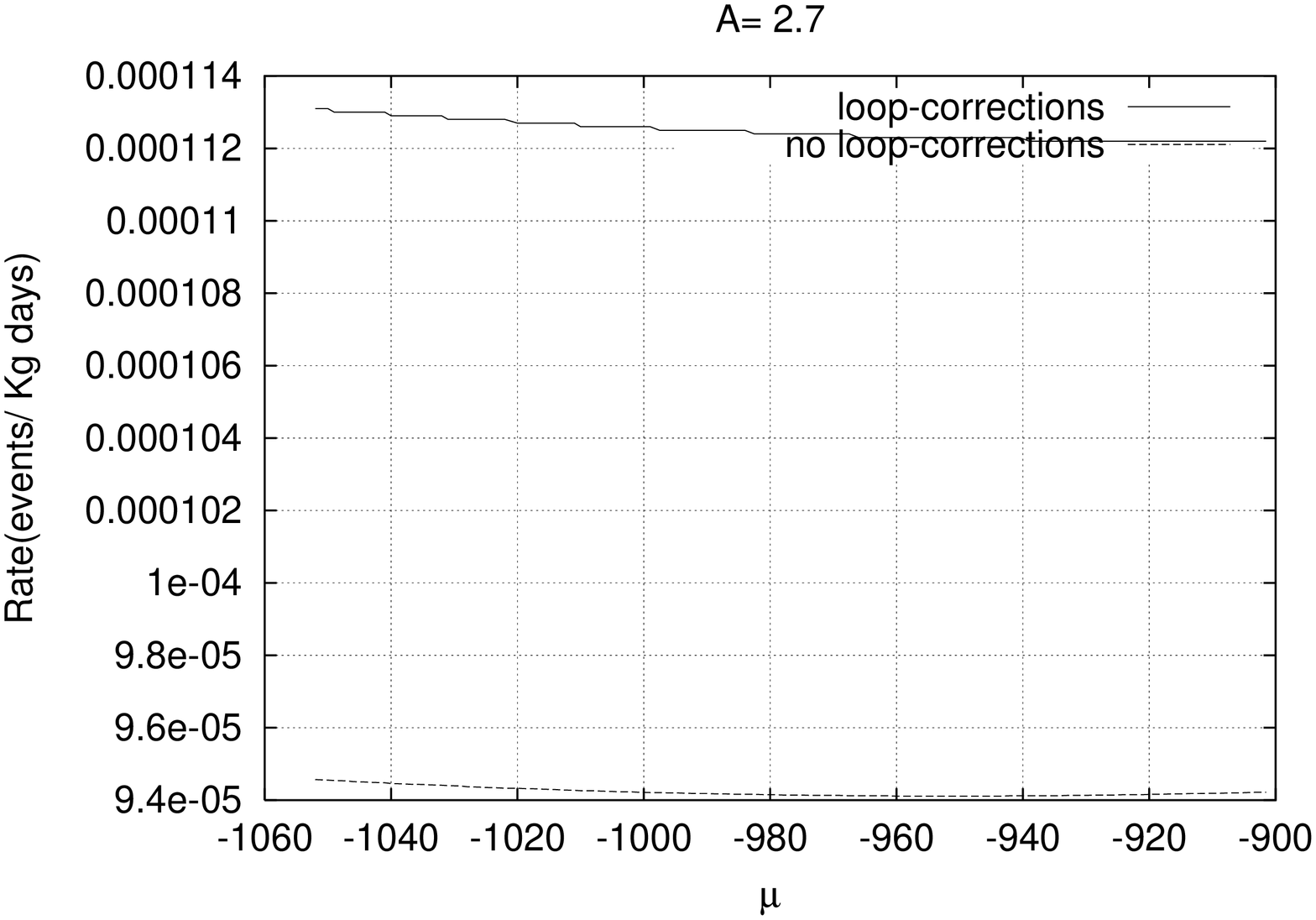,angle=-90,width=0.60\textwidth}}
\caption{}
\label{f13}
\end{figure}

\begin{figure}
\centerline{\psfig{file=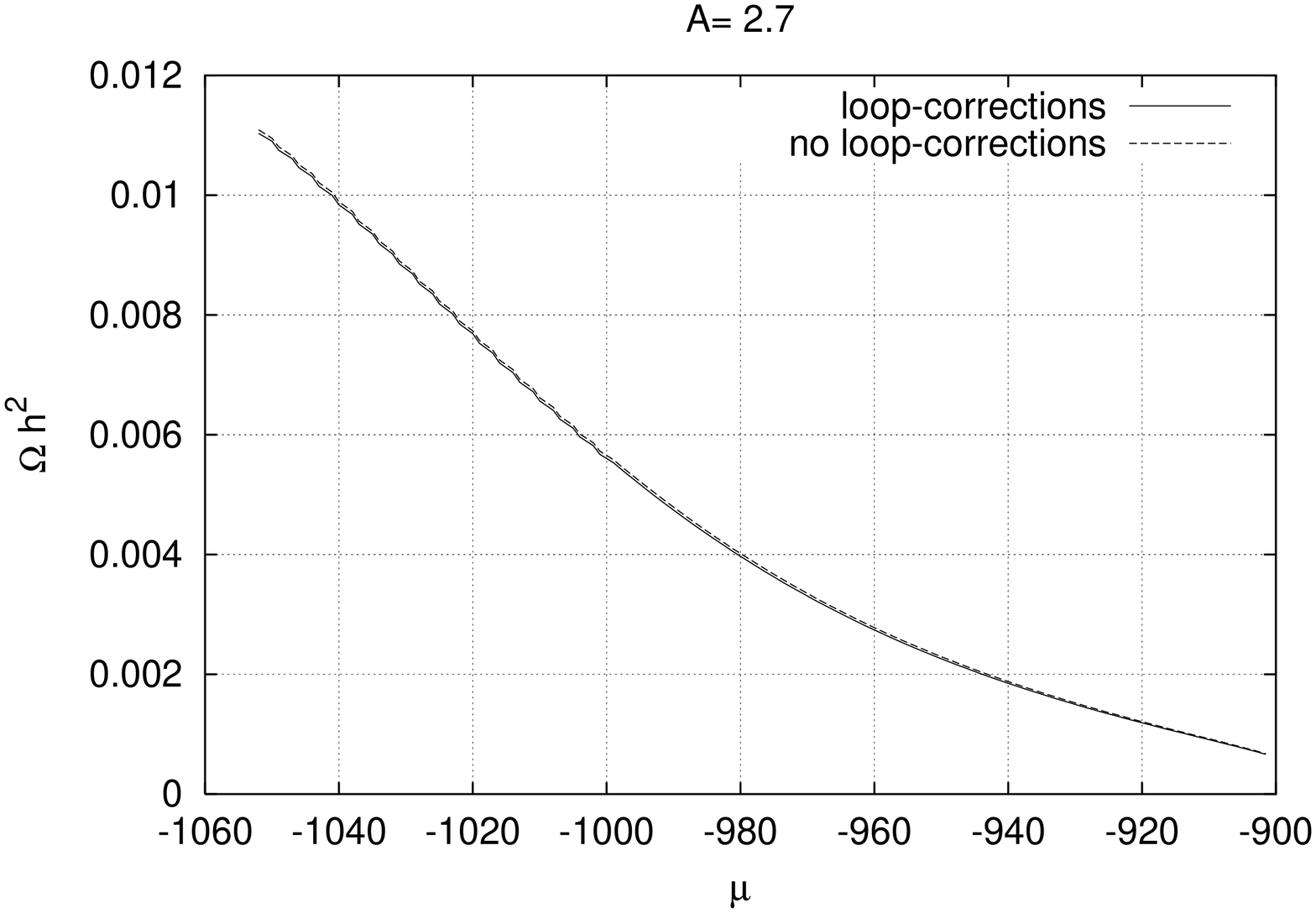,angle=-90,width=0.60\textwidth}}
\caption{}
\label{f14}
\end{figure}

\begin{figure}
\centerline{\psfig{file=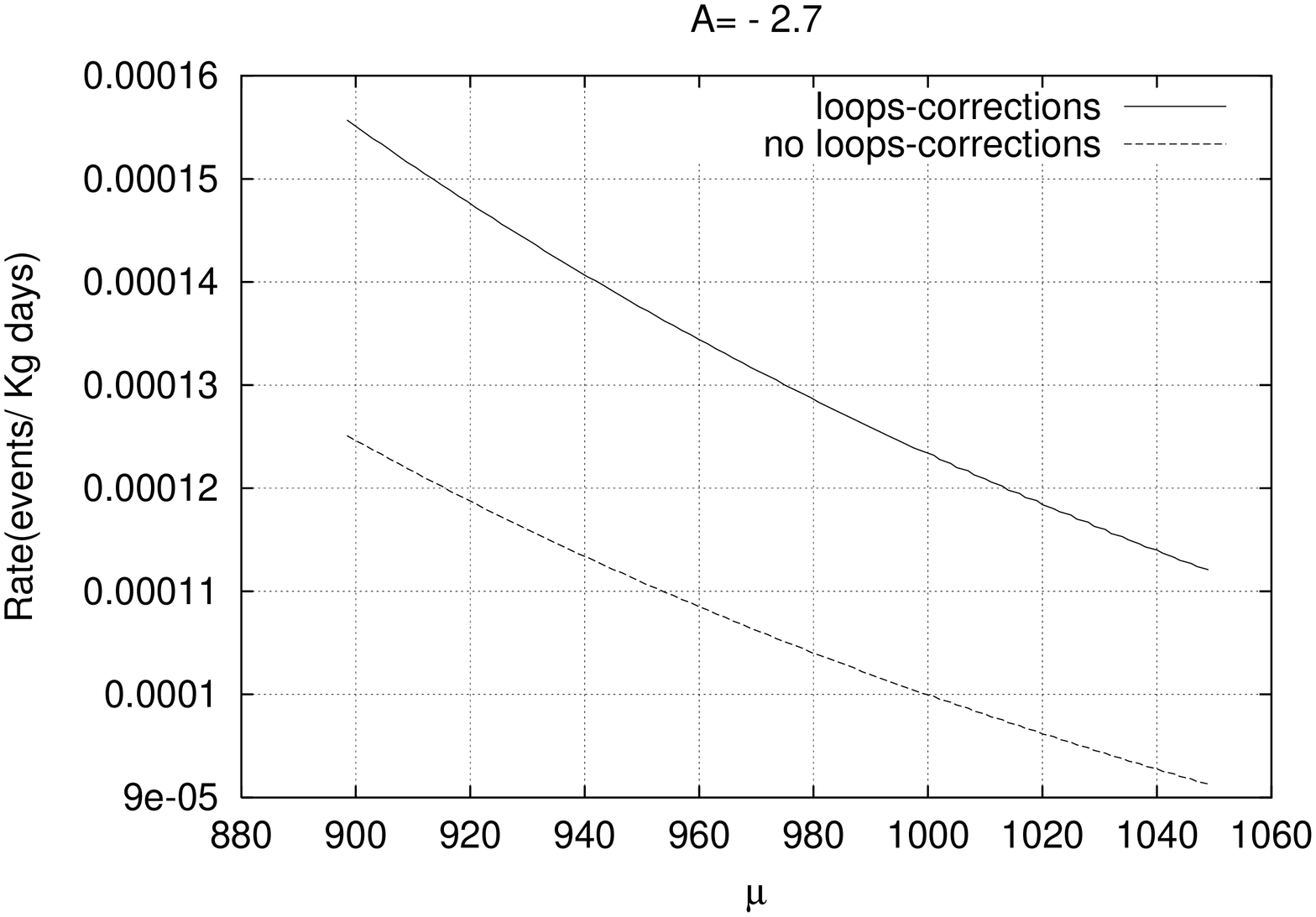,angle=-90,width=0.60\textwidth}}
\caption{}
\label{f15}
\end{figure}

\begin{figure}
\centerline{\psfig{file=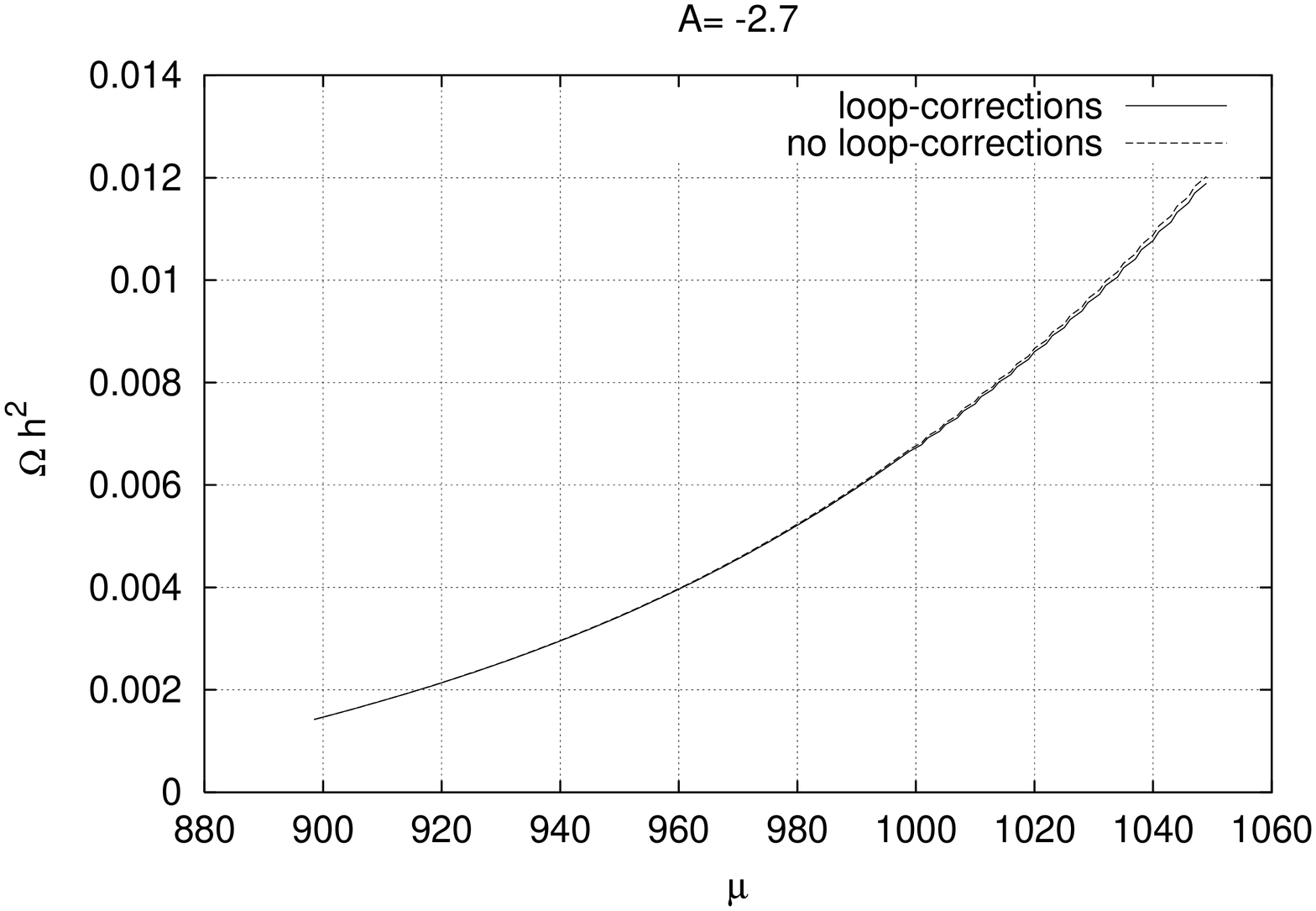,angle=-90,width=0.60\textwidth}}
\caption{}
\label{f16}
\end{figure}

In our case, we are interested in the Bino limit, 
where $M_1 < M_2 , \mu$. In the soft breaking term, we will consider $A_t = A_b =A$ and 
$M_{\widetilde{Q}}=m_{\widetilde{t_R}}=m_{\widetilde{b_R}}$ for simplicity.  
\\
Knowning the details related with the counting rate and the experimental contraints for SUSY particles, we will show 
some numerical examples, where we compute the counting rate including and excluding the new loop corrections 
to the LSP-LSP-Higgs boson coupling.
\\
For these calculations, we will use: $M_{\widetilde{Q}}= 400 \ \rm{GeV}$, $m_{A^0}= 500 \ \rm{GeV}$, the Bino mass $M_1 = 200 \ \rm{GeV}$, 
the Wino mass $M_2 = 400 \ \rm{GeV}$ and $\tan \beta = 5$. The parameter $\mu$ is varied from $\pm 500 \  \ \rm{GeV}$ to $\pm 1050 \ \ \rm{GeV}$ in 
general. For $A$ we consider the values $\pm 2.5 $, $\pm 2.6$ and $\pm 2.7$ $M_{\widetilde{Q}}$.
\\
Note that in this case, the mass spectrum for supersymmetric particles is above the SM particles mass spectrum. It 
means that in this case we have not overlapping of these mass spectrums and all sparticle mass bounds are easily satisfied.
\\
The first example is shown in Fig \ref{f4} and Fig \ref{f5}. In this case we show the values for the rate and the relic 
density respectively, when $A=2.5$ and $-1050 \leq \mu \leq -500$. The counting rate is incremented when the loop-corrections are 
considered, where the bigger difference is of a 20 \% in the interval $-1050 \leq \mu \leq - 900$.  Analyzing the relic 
density, we have a significant value in the interval $-1050 \leq \mu \leq -950$, and when the loop-corrections are 
included, it does not change appreciably. Note that the mass of the lighter stop squark increases with increasing $\mu$. 
The upper bound on $\mu$ is set by the requirement that the neutralino must be the lightest SUSY particle. For the upper 
bound, the relic density is greatly reduced by $\widetilde{\chi ^0_1}- \widetilde{t_1}$ co-annihilation \cite{Drees}.    
\\
For the opposite choise of signs $A= - 2.5$ and $500 \leq \mu \leq 1050$ (see Fig \ref{f6} and Fig \ref{f7}), the counting rate does not change. The 
relic density for these values has a significant value in the interval $ 900 \leq \mu \leq 1050$ and it does not change, 
when the loop corrections are considered. Note that now $m_{\widetilde{t_1}}$ increases with increasing $\mu$ and the effect 
of $ \widetilde{t_1}- \widetilde{\chi^0_1}$ co-annihilation is again evident.
\\
In Fig \ref{f9} and Fig \ref{f10}, the counting rate and the relic density for $A=2.6$ and $-1050 \leq \mu \leq -700$ are shown.
The counting rate is incremented by a  16-20 \% if the loop-corrections are 
included, while the relic density is quite small in this interval, equal to $\Omega h^2 \simeq 0.1$ for $-1050 \leq \mu \leq -1000$.
\\
Considering the loop-corrections, in the calculation of $R$ and $ \Omega h^2$ for $A= - 2.6$ and 
$ 700 \leq \mu \leq 1050$ (Fig \ref{f11} and Fig \ref{f12}), we see 
that $R$ is increased and $\Omega h^2$ remain equal, but it is very small in this case.
\\
In the last example, we consider $\vert A \vert = 2.7$ and $900 \leq \vert \mu \vert \leq 1050$, 
taking into account two different cases for the different signs of 
these parameters. When $A= 2.7$ and $\mu$ is varied from -1050 to -900 (Fig \ref{f13} and Fig \ref{f14}),
the inclusion of loop corrections increases $R$, while the relic density is 
very small in this interval. Changing the sign of these parameters(Figs \ref{f15} and \ref{f16}), 
$R$ is increased and $\Omega h^2$ remains equal.
\\
We see that for the different values and signs considered of $A$ and $\mu$ 
the Rate is increased, but the size of the correction is modest $ ( \simeq 20 \% )$.
\\
The size of the corrections increases with decreasing $m_{\widetilde{t_1}}$. However, one has to respect 
the constraint $m_{ \widetilde{t_1} } > m_{\widetilde{\chi^0_1}}$. Moreover, 
for $m_{\widetilde{t_1}} \simeq m_{\widetilde{\chi^0_1}}$ the predicted LSP relic density is small, due to co-annihilation. It should be noted, that in our calculation only include thermal relics, which were in thermal equilibrium after inflation.
There might be significant contribution to $ \Omega h^2 $, from non-thermal relics \cite{Randall} 
\\
In this way, we conclude that for the predictions of new experiments to search for SUSY Dark Matter candidates, might eventually be necessary 
to include loop corrections to the counting rate. However, these corrections are only important when the 
gaugino-higgsino mixing is small, i.e when the LSP is nearly pure Bino. In this case the predicted counting rate is unfortunately several orders of magnitude below the sensitivity of present experiments. 

\chapter{Conclusions.}

The different Non-Baryonic Dark Matter particle candidates 
and the basic elements of the Minimal Supersymmetric 
Extension of the Standard Model needed for our purposes 
were reviewed.
\\
To improve the calculation of the neutralino counting rate were introduced 
new Yukawa loop corrections to the coupling 
neutralino-neutralino-Higgs boson in the Bino-limit.
\\
We have shown several examples to analyze the contributions 
of these corrections on the counting rate, considering 
the neutralino mass $M_1 = 200 \ \rm{GeV}$. The difference 
in the neutralino counting rate when the loop corrections 
are considered can reach 20\%. Since these corrections 
are significant only when the counting rate is small, it is not important 
for present or near-future experiments. However, 
it plays an important role when one is trying to 
determine the minimal realistic counting rate in the MSSM \cite{Gondolo}.  
\\
The interesting interval for the relic density was 
taken into account to define when the loop corrections 
are important, the supersymmetric parameters were 
chosen according to the experiment constraints given at LEP2 \cite{Data}.
\\
In general we conclude that it will be interesting to 
consider these Yukawa one-loop corrections in the different 
limits of neutralino physical states, 
and study the intervals of the different supersymmetric parameters 
which would provide the possibility of improving the counting rate 
appreciatly.
\\

\newpage

\textbf{{\LARGE Acknowledgments.}}
\\
\\
I would like to thank the Abdus Salam International Centre for Theoretical Physics (ICTP), 
for providing me the possibility of studying in the Diploma Course, the High Energy Physics 
Group Staff for the excellent lectures on the different 
topics, and the help to understand the basic elements to start in High Energy Physics.
\\
I would like to thank my supervisor Professor Manuel Drees for his great 
support during the development of this dissertation.
\\
I also want to thank Ms Concetta Mosca for her great help during my stay at the 
ICTP and collaboration in the different things related with our lifes at the Centre.
\\
I thank the Max Planck Institut f\"ur Physics (Werner-Heisenberg-Institut), 
where I am starting my Ph.D in High Energy Physics.

\bibliographystyle{plain}

\newpage
\pagestyle{empty}

\textbf{{\LARGE APPENDIX A.}}
\\

The loop integrals \cite{loop}:

\begin{displaymath}
C_0 (p_{1}, p_{2}, M_{1}, M_{2}, M_{3}) = (2 \pi \mu)^{4-n} \int \frac{d^{n} l}{i \pi^2} \frac{1}{D_{1}D_{2}D_{3}}
\end{displaymath}

with

\begin{displaymath}
D_1 = l^{2} - M^2_1 + i \epsilon
\end{displaymath}

\begin{displaymath}
D_2 = (l + p_{1})^{2} - M^2_2 + i \epsilon
\end{displaymath}

\begin{displaymath}
D_3 = (l + p_{1}+p_{2})^{2} - M^2_3 + i \epsilon     
\end{displaymath}

\begin{displaymath}
\overline{C}_{\mu} (p_{1},p_{2}, M_{1}, M_{2}, M_{3}) = 
(2 \pi \mu)^{4-n} \int \frac{d^{n}l}{i \pi^2} \frac{l_{\mu}}{D_{4}D_{5}D_{6}}
\end{displaymath}

where:

\begin{displaymath}
D_4 = (l - p_{1})^{2} - M^2_1 + i \epsilon
\end{displaymath}

\begin{displaymath}
D_5 = (l - p_{2})^{2} - M^2_3  + i \epsilon
\end{displaymath}

\begin{displaymath}
D_6 = l^{2} - M^2_1 + i \epsilon
\end{displaymath}

\begin{displaymath}
\overline{C}_{\mu}=C^{+}_{1}(p_1+p_2)_{\mu}+C^{-}_{1}(p_1 - p_2)_{\mu }
\end{displaymath}

\newpage
\pagestyle{empty}
\begin{center}
\underline{ \textbf{{\LARGE FIGURES.}}}
\end{center}

\newpage

\end{document}